\documentclass[twocolumn,english,pre,floatfix,citeautoscript,nofootinbib]{revtex4}
\usepackage{amsbsy}
\usepackage{latexsym,epsfig,graphicx}
\usepackage{dcolumn}
\usepackage{subfigure}
\usepackage{comment}
\usepackage{color}
\usepackage[colorlinks,urlcolor=blue,citecolor=blue]{hyperref}
\usepackage{amstext}
\usepackage{amssymb}
\usepackage{setspace}
\usepackage{amsmath}
\usepackage{makeidx}
\usepackage{bm}
\usepackage{ulem}
\usepackage{multirow}
\usepackage{mathrsfs}

\usepackage{array}

\newcommand{\ii}{\mathrm{i}}
\newcommand{\e}{\mathrm{e}}
\newcommand{\dd}{\mathrm{d}}

\newcommand{\diag}{\mathrm{diag}}

\def\Im{\textup{Im}}
\def\Re{\textup{Re}}

\begin{document}
\title{Multihump-Multivalley Soliton Families on a Plane Wave Background in Birefringent Optical Fibers}
\author{Jin-Peng Yang$^{1}$}
\author{Yan-Hong Qin$^{2}$}\email{yhqin@xju.edu.cn}
\address{$^{1}$College of Mathematics and System Sciences, Xinjiang University, Urumqi 830046, China}
\address{$^{2}$Institute of Mathematics and Physics, Xinjiang University, Urumqi 830046, China}

\begin{abstract}

We obtain a family of multihump-multivalley solitons (MHMVSs) on a plane-wave background in birefringent optical fibers governed by the two-component Fokas-Lenells equations, with exact solutions derived via the Darboux transformation method. The fundamental solutions are systematically classified through their phase diagrams, and higher-order configurations are identified as well. Notably, the construction extends to solitons with arbitrary MHMV structures, a class of solutions previously unreported in two-component integrable systems. Numerical simulations confirm the robustness of these solutions under weak white noise. Furthermore, analysis of their topological structure reveals that the virtual monopole field is determined equally by the intensity zeros and poles of MHMVSs in the complex plane, a feature that has remained unrecognized in previous studies of nonlinear wave topological phases. These findings reveal a previously unknown soliton family on a plane-wave background along with a distinct topological feature within the two-component framework, thereby enriching the broader understanding of topological phases of nonlinear waves.

\end{abstract}
\date{\today}
	
\maketitle
	
\section{Introduction}

Vector solitons have been extensively investigated experimentally and theoretically in many nonlinear systems, particularly in Bose-Einstein condensates and nonlinear optical fiber systems owing to their high experimental accessibility \cite{OS1,OS2,OS3,OS4,BEC1,BEC2,BEC3}. Compared to scalar solitons, they exhibit richer dynamics with distinct configurations such as bright-bright \cite{BB}, dark-dark, dark-bright \cite{DB1,DB2}, dark-antidark \cite{DAD1}, and antidark-antidark solitons \cite{DAD2}. Extensive research has been devoted to obtaining exact analytical solutions for these nonlinear waves in the nonlinear Schr\"{o}dinger equation hierarchy \cite{DB3,Ling3,mao}, which are valued for both their mathematical elegance and ability to elucidate fundamental physics.

In recent years, research on exploring analytical solutions for novel vector solitons in coupled nonlinear systems has gained increasing attention due to the abundant nonlinear interactions between components, yielding promising results such as magnetic solitons (a new type of antidark-dark solitons) \cite{MS-Qu,MS-Chai,MS-Chai2}, spin solitons (a new class of  bright-dark solitons) \cite{SS-zhao,SS-zhao2,SS-zhao3}, nondegenerate solitons \cite{NS-Stalin1,Qin1,NS-Stalin2,Qin2,zhao3,NS-Lin,Qin}, multivalley dark solitons \cite{Qin2,zhao3,DV-ling,DV-ling1,Qin3}, and dark-bright solitons with striking width differences \cite{mao1}. These solutions have been obtained by developing various mathematical methods such as the Darboux transformation (DT) method, the Hirota bilinear method, the decoupling transformation method, and variational Lagrangian approaches, among other techniques. These novel vector solitons exhibit unique dynamical behaviors with potential applications. For instance, spin solitons demonstrate AC oscillations under a constant force, enabling weak force detection \cite{SS-zhao,SS-zhao2}; magnetic solitons provide a platform for studying magnetic excitation mechanisms \cite{MS-Qu,MS-Chai,MS-Chai2}; nondegenerate solitons enable the construction of complete eigenstate sets in $N$-multiple well potential systems \cite{Qin1,Qin2,zhao3,zhao4}. These findings suggest that investigating novel nonlinear wave solutions could serve as a critical avenue for unveiling the underlying physics.

Multihump and multivalley solitons account for a significant share of research in this field, particularly focusing on nondegenerate solitons and multivalley dark solitons on plane wave backgrounds in various physical systems. In the two-component Manakov system, nondegenerate solitons were first obtained using  Hirota bilinear method \cite{NS-Stalin1,NS-Stalin2} and  DT \cite{Qin1,Qin2}. Subsequently, they have also been studied in various two-component systems, including Newell-type long-wave-short-wave models \cite{NS-He}, the long-wave-short-wave resonance interaction system \cite{NS-Stalin3}, mixed derivative nonlinear Schr\"{o}dinger equations \cite{Dai1} and physics-informed neural networks \cite{Jaganathana}. These nondegenerate solitons are primarily associated with multihump bright solitons on a vanishing background in all components, or multihump bright solitons in some components coupled to multivalley dark solitons in others.  The double-valley dark soliton can be derived in three-component Manakov systems \cite{Qin2,zhao3} and scalar defoucsing Hirota systems \cite{DV-ling}. It also points out that the $(N-1)$-valley dark soliton can be derived in $N$-component Manakov systems \cite{Qin2}. These multivalley dark solitons demonstrate multiple phase steps and striking collision dynamics \cite{Qin2,phase-zhao}. Dark-antidark soliton arrays were recently observed in two-component Bose-Einstein condensates with varying interspecies interaction strengths \cite{Katsimiga}. However, to the best of our knowledge, the multihump antidark soliton or multihump and multivalley complexes on a plane wave background remain analytically unexplored.

This work reports the first analytical demonstration of MHMVSs on a plane wave background in birefringent fibers governed by the coupled Fokas-Lenells (CFL) equations.  Using the DT, we obtain exact solutions for such solitons, ranging from fundamental to higher-order and even arbitrary configurations. The fundamental solutions exhibit diverse intensity profiles, such as double-valley dark/double-hump antidark, anti-W-shaped/W-shaped, and dark–antidark complexes/dark–antidark complexes (see examples in Fig.~\ref{fig1}), where the slash (/) separates the two components of a vector soliton. Phase diagrams for a systematic classification of fundamental MHMVSs have been established based on intensity extrema.  Beyond these fundamental patterns, higher-order configurations including double-hump-double-valley/double-hump-double-valley and double-hump-triple-valley/triple-hump-double-valley patterns are also identified (see examples in Fig.~\ref{fig3}).  Moreover, via the topological vector potential theory, we demonstrate the multiple-step phase shift of MHMVS. We further reveal that its virtual monopole field is composed of both zeros and poles of the intensity in the complex plane. Particularly, half of the virtual monopoles originate from zeros, and the other half from poles. This equal partition has not been reported in studies of topological phases of nonlinear waves. Furthermore, a key finding is that an $N$-fold DT with appropriate constraints can generate static arbitrary MHMVS (AMHMVS) configurations in a two-component system, including $K$-hump antidark solitons, $M$-valley dark solitons (where $K$ and $M$ are arbitrary positive integers), and their hybrid complexes (see examples in Fig.~\ref{fig6}), which have not been realized in two-component nonlinear Schr\"{o}dinger systems before. Numerical simulations under weak white noise perturbations confirm the dynamical stability of both MHMVS and AMHMVS. These novel solitons arise from the coherent superposition of solitons within the same component, in contrast to nondegenerate solitons, which emerge from the incoherent superposition of solitons across different components \cite{NS-Stalin1,Qin1,NS-Stalin2,Qin2,zhao3,NS-Lin,Qin}.

The remainder of this paper is organized as follows.  In Sec.~\ref{sec2}, the fundamental MHMVSs solutions of the CFL equations are constructed via the DT by imposing the velocity constraint and the background amplitude constraint on the multi-dark/antidark soliton solution. We classify their intensity profiles based on intensity extrema and establish the corresponding phase diagrams in the appropriate parameter space.  Furthermore, we derive the AMHMVSs and demonstrate that they exist only in the static case.  In Sec.~\ref{sec3}, we analyze the multiple-step phase shift and reveal, via topological vector potential theory, that the virtual monopole field is equally contributed by intensity zeros and poles.  Sec.~\ref{sec4} provides numerical simulations under weak white noise perturbations, examining the dynamical stability of both MHMVSs and AMHMVSs. Conclusions and discussions are given in Sec.~\ref{sec5}.

\section{Multihump-multivalley soliton solutions to the coupled Fokas-Lenells equation on a plane wave background}\label{sec2}

\subsection{Physical model and multihump-multivalley soliton solutions} \label{sec2A}

\begin{figure*}[!t]
	\centering		
	\includegraphics[width=1.95\columnwidth]{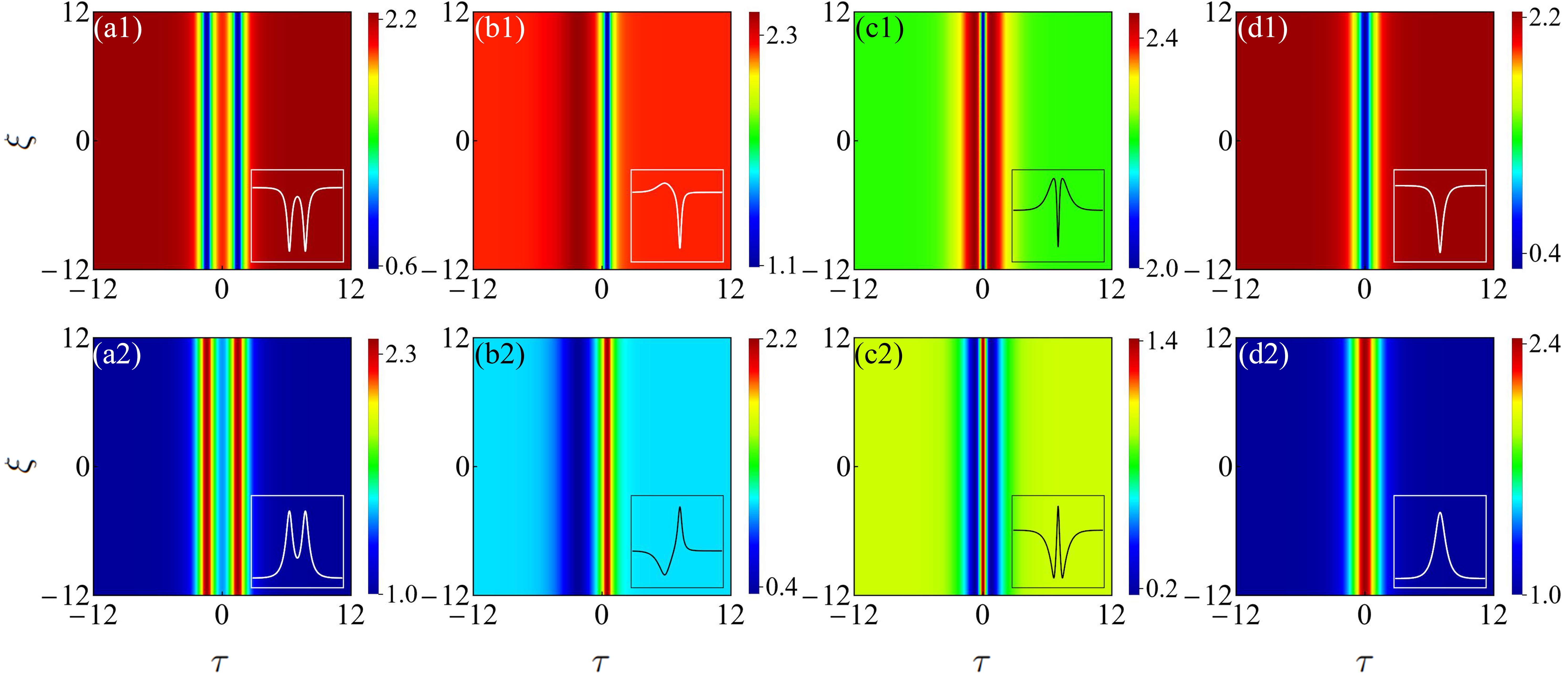}
	\caption{Some intensity structures of MHMVSs on the plane wave. First row corresponds to the first component, and second row to the second component. The points ($\varepsilon_1$,$\varepsilon_2$) are chosen as follows: (a): $\varepsilon_1=\varepsilon_2= -\frac{1}{\sqrt{5}}$, corresponding to the cube pattern; (b):  $\varepsilon_1=2,\varepsilon_2= -0.5$, corresponding to the star pattern; (c): $\varepsilon_1=-\varepsilon_2= \frac{1}{\sqrt{5}}$, corresponding to the regular hexagonal pattern;  (d): $\varepsilon_1=-\varepsilon_2= -\frac{1}{\sqrt{5}}$, corresponding to the triangular pattern. {Other parameters are selected as follows:  $b_1=-b_2=0.5,z_1=0.5,z_2=1,a_1=\sqrt{5},a_2=1,\kappa_1=0.5+0.5\ii,\kappa_2=1+0.5\ii$.} }
	\label{fig1}
\end{figure*}

Ultrashort optical pulse transmission in birefringent fibers can be described by a system of CFL equations \cite{Chen1,Ye1}. For our analysis of pulse propagation, we consider the normalized CFL form:
\small
\begin{subequations}\label{0}
\begin{align*}
&\ii D_\tau \frac{\partial u_1}{\partial \xi}\!-\!\frac{\eta}{2}\frac{\partial^2 u_1}{\partial \tau^2}\!+\!\ii(2|u_1|^2\!+\!|u_2|^2)D_\tau u_{1}\!+\!u_1u_2^*D_\tau u_{2}\!=\!0,\\
&\ii D_\tau \frac{\partial u_2}{\partial \xi}\!-\!\frac{\eta}{2}\frac{\partial^2 u_2}{\partial \tau^2}\!+\!\ii(2|u_2|^2\!+\!|u_1|^2)D_\tau u_{2}\!+\! u_2u_1^*D_\tau u_{1}\!=\!0,
\end{align*}
\end{subequations}
\noindent where $u_1(\xi,\tau)$ and $u_2(\xi,\tau)$ represent the complex envelopes of the two field components, with $\xi$ denoting the propagation distance and $\tau$ the retarded time. The subscripts indicate partial derivatives, and the asterisk denotes complex conjugation. The parameter $\eta=\pm 1$ denotes the type of dispersion, i.e.,  $-1$ for anomalous dispersion and  $+1$ for normal dispersion. The differential operator $D_\tau = 1 + \mathrm{i}\varrho \partial_\tau$ operates on $\tau$, where $\varrho$ scales the perturbation to the Manakov system (the $\varrho=0$ case). The term $\mathrm{i}\varrho \partial_\tau$ in $D_\tau \partial_\xi u_1$ and $D_\tau \partial_\xi u_2$ represents the space-time coupling effect. $D_\tau u_1$ and $D_\tau u_2$ in the last two terms accounts for self-steepening effects. These additional terms correct the slowly varying envelope approximation when modeling few-cycle pulse propagation---a regime encountered in both ultrafast optics and hydrodynamic wave systems. This makes the CFL equations particularly suitable for describing femtosecond pulse dynamics in birefringent fibers and ocean wave dynamics. As an integrable extension of the Manakov model, some localized waves of CFL equations \eqref{0} have been obtained \cite{Chen1,Ye1,Ling1,su,Ling2}, such as rogue waves and vector solitons. To facilitate the analysis, we introduce the following transformation \cite{su}
$$\tau\mapsto \varrho\eta(\tau-\xi),\xi\mapsto-2\varrho^2\xi,u_s\mapsto \frac{\ii}{2\varrho}\e^{\eta\ii(\tau+\xi)}u_s ~~(s=1,2).$$
Then system (\ref{0}) yields the following CFL equations 
\begin{subequations}\label{1}
	\begin{align}
		&u_{1,\tau\xi}\!+\! u_1\!+\!\ii(|u_1|^2\!+\!\frac{1}{2}|u_2|^2)u_{1,\tau}\!+\!\frac{\ii}{2} u_1u_2^*u_{2,\tau}\!=\!0,\\
		&u_{2,\tau\xi}\!+\! u_2\!+\!\ii(\frac{1}{2}|u_1|^2\!+\!|u_2|^2)u_{2,\tau}\!+\!\frac{\ii}{2} u_2u_1^*u_{1,\tau}\!=\!0.
	\end{align}
\end{subequations}
The CFL system \eqref{1} is distinguished from the standard Manakov model by its intricate self-steepening and spatio-temporal coupling effects. For our present purpose of investigating soliton structures, its most crucial feature is that this system admits two fundamental types of vector solitons on a plane wave background: dark solitons (DS) and antidark solitons (ADS) \cite{Ling1}. In previous studies, analytical investigations of multihump solitons have been largely confined to the vanishing background, where nondegenerate bright solitons can exhibit double-hump or even multihump profiles \cite{NS-Stalin1,Qin1,NS-Stalin2,Qin2,zhao3,NS-Lin,Qin}. On a plane wave background, by contrast, existing studies have primarily focused on multivalley dark solitons in nonlinear Schr\"{o}dinger systems \cite{Qin2,Qin3,zhao3,DV-ling,DV-ling1}. However, multihump antidark solitons on a plane wave background have remained analytically unexplored. To the best of our knowledge, a systematic analytical study encompassing multihump, multivalley, and their hybrid structures on a plane wave background is still absent from the literature. The CFL system provides a natural platform to fill this gap.

To carry out this investigation, we revisit the general multi-antidark/dark soliton solutions of Eq.~\eqref{1} \cite{Ling1}, obtained via the DT from the plane wave solution $u_s=a_s\e^{\ii \omega_s},$ 
where
$\omega_1=b_1\tau+\frac{1}{2}\big[\frac{2}{b_1}-(2|a_1|^2+|a_2|^2)-|a_2|^2\frac{b_2}{b_1}\big]\xi$, $
\omega_2=b_2\tau+\frac{1}{2}\big[\frac{2}{b_2}-(|a_1|^2+2|a_2|^2)-|a_1|^2\frac{b_1}{b_2}\big]\xi$,
$a_s,~b_s\in\mathbb{R}\backslash\{0\}$. The parameters $a_s$ and $b_s$ represent the background amplitude and wavenumber, respectively. In this work we consider $b_1\neq b_2$. The multi-antidark/dark solitons can be expressed as \cite{Ling1}
\begin{subequations}\label{solution}
\begin{align}
u_1[N]&= a_1\frac{~~\det H^{[1]}}{\det H}\e^{\ii\omega_1},\\
u_2[N]&= a_2\frac{~~\det H^{[2]}}{\det H}\e^{\ii\omega_2},
	\end{align}
\end{subequations}
with
\begin{align*}
H_{ij}&=\frac{\kappa_{i}^*}{\kappa_{j}-\kappa_{i}^*}\e^{\vartheta_{i}^*+\vartheta_{j}}+\frac{\varepsilon_i\delta_{ij}|\kappa_{i}|}{\kappa_{i}-\kappa_{i}^*}, 1\le i,j\le N,\\	H^{[1]}_{ij}&=\frac{\kappa_{j}}{\kappa_{j}-\kappa_{i}^*}\frac{b_1+\kappa_{i}^*}{b_1+\kappa_{j}}\e^{\vartheta_{i}^*+\vartheta_{j}}+\frac{\varepsilon_i\delta_{ij}|\kappa_{i}|}{\kappa_{i}-\kappa_{i}^*},\\
H^{[2]}_{ij}&=\frac{\kappa_{j}}{\kappa_{j}-\kappa_{i}^*}\frac{b_2+\kappa_{i}^*}{b_2+\kappa_{j}}\e^{\vartheta_{i}^*+\vartheta_{j}}+\frac{\varepsilon_i\delta_{ij}|\kappa_{i}|}{\kappa_{i}-\kappa_{i}^*},\\
\vartheta_i&=- \Im(\kappa_i)(\tau-v_i\xi)+\ii\left[(\Re(\kappa_i)-z_i)\tau+t_i\xi\right].
\end{align*}
Here $\delta_{ij}$ is the Kronecker delta and $\varepsilon_i$ is a real parameter controlling the soliton's spatial position.
$v_i=-\frac{1}{2b_1b_2z_i}[2\Re(\kappa_{i})-2z_i+b_1+b_2]$ is the velocity of the single soliton, $t_{i}=\frac{1}{4z_i}-\frac{z_i}{2b_1b_2}+\frac{b_1+b_2}{b_1b_2}(a_1^2b_1+a_2^2b_2-1)-[\Re(\kappa_i)-z_i]v_i-\frac{|\kappa_i-z_i|^2}{2b_1b_2z_i}$ is real.  $z\in \mathbb{R}\backslash\{0\}$ is the spectral parameter.  $\kappa_i$ is a solution of the spectral equation
\begin{align}\label{spectrumcondition}
	\frac{\kappa}{z}-2+\frac{a_1^2b_1^2}{\kappa+b_1}+\frac{a_2^2b_2^2}{\kappa+b_2}=0
\end{align}
as $z=z_i$, satisfying $\Im(\kappa_i)\neq 0$.
By examining the central intensity of the single soliton, one finds that $b_s\varepsilon_1>0$ and $b_s\varepsilon_1<0$ correspond to an ADS and a DS, respectively. Due to the existence of both DS and ADS, nonlinear superposition enables the construction of solitons featuring multihump, multivalley, as well as hybrid structures combining both humps and valleys on a plane wave background.  We shall collectively refer to such solitons as MHMVSs. Their construction, however, requires more than just the coexistence of DS and ADS.

It has been established in previous studies that, in vector systems, the nonlinear superposition of multisolitons to form multihump or multivalley structures requires a necessary condition: the participating solitons must have the same velocity \cite{DV-ling1,Qin1,Qin2,NS-Stalin2,NS-Stalin1,NS-Lin,zhao3,DV-ling,Qin}. Consequently, the velocity of solitons plays a particularly crucial role.  However, the velocity constraint alone does not suffice to construct such solitons. As previous studies have demonstrated, the construction of nondegenerate solitons necessitates additional eigenfunction constraints \cite{NS-Stalin1,NS-Stalin2,Qin1,Qin,NS-Lin}, while solitons on a plane wave background require spectral constraints to achieve multisoliton superpositions \cite{DV-ling,DV-ling1}. Guided by these findings, the first step in our construction of MHMVSs is to adopt the velocity as an independent control parameter so as to realize the velocity resonance condition. To this end, we express $\kappa$ as a function of $z$ and $v$. From the velocity expression, one directly obtains
\begin{align}\label{14}
	\Re(\kappa)&=z(1-b_1b_2v)-\frac{b_1+b_2}{2}.
\end{align}
Then, the spectral constraint is employed to derive an additional constraint. Although the spectral equation \eqref{spectrumcondition} does not explicitly involve the velocity, it can be reformulated by applying Cramer's rule to the spectral condition and its conjugate, yielding the background amplitude functions $a_s^2(z,\kappa)$. Under this reformulation, the velocity enters through the relation \eqref{14}. For two distinct spectral parameters $z_p,z_q$, the DT requires that the background amplitudes remain equal: $a_s^2(z_p,\kappa_p)=a_s^2(z_q,\kappa_q)$. This background amplitude constraint, together with the velocity one, constitutes the two necessary constraints that determine admissible $\kappa_p,\kappa_q$ for constructing MHMVSs. For $v_p=v_q\neq 0$, background amplitude condition gives the unique solution 
\begin{subequations}\label{16}
	\begin{align}
		[\Im(\kappa_p)]^2=&(1-b_1b_2v_p)[(b_1+b_2)z_p+4b_1b_2v_pz_pz_q]\nonumber\\
		&-(\Delta b)^2-z_p^2(1-b_1b_2v_p)^2,\\
		[\Im(\kappa_q)]^2=&(1-b_1b_2v_p)[(b_1+b_2)z_q+4b_1b_2v_pz_pz_q]\nonumber\\
		&-(\Delta b)^2-z_q^2(1-b_1b_2v_p)^2.
	\end{align}
\end{subequations}
where $\Delta b=\frac{b_1-b_2}{2}$. When velocity condition $v_p=v_q=0$, one can obtain
\begin{align}\label{18}
	[\Im(\kappa_p)]^2z_p^{-1}=[\Im(\kappa_q)]^2z_q^{-1}+f,
\end{align}
where $f=z_q-z_p+(\Delta b)^2(z_q^{-1}-z_p^{-1}).$  See Appendix \ref{App_k}  for the derivation of Eqs. (\ref{16}) and (\ref{18}). Eqs. (\ref{14}) and (\ref{16}) determine a non-static MHMVS corresponding to $z_p$ and $z_q$, while Eqs. (\ref{14}) and (\ref{18}) determine a static one.

In general, by taking $N=2$ in solution (\ref{solution}) with background amplitude constraint and the velocity constraint, and introducing the shorthand notation
$\rho=\left|\frac{\kappa_2-\kappa_1}{\kappa_2^*-\kappa_1}\right|^2$, ${\kappa}_j^{[s]}=\frac{\kappa_j(b_s+\kappa_j^*)}{b_s+\kappa_j}$, $\theta_j=\Im(\kappa_j)(\tau-v_1\xi)$ ($j=1,2$ and $v_1=v_2$),
the expression of MHMVS can be given by
\begin{align}\label{b_mhmv_sol}
	u_s[2]
	&= a_s\frac{\Lambda^{[s]}}{\Gamma}\e^{\ii\omega_s},
\end{align}
where
\begin{align*}
	\Lambda^{[s]}=&\rho{\kappa}_1^{[s]}{\kappa}_2^{[s]}\e^{-\theta_1-\theta_2}+{\varepsilon_1\varepsilon_2|\kappa_1\kappa_2|}\e^{\theta_1+\theta_2}\\
	&+{\varepsilon_1|\kappa_1|}{\kappa}_2^{[s]}\e^{\theta_1-\theta_2}+{\varepsilon_2|\kappa_2|}{\kappa}_1^{[s]}\e^{\theta_2-\theta_1},\\
	\Gamma
	=&\rho{\kappa}_1^{*}{\kappa}_2^{*}\e^{-\theta_1-\theta_2}+{\varepsilon_1\varepsilon_2|\kappa_1\kappa_2|}\e^{\theta_1+\theta_2}\\
	&+{\varepsilon_1|\kappa_1|}{\kappa}_2^{*}\e^{\theta_1-\theta_2}+{\varepsilon_2|\kappa_2|}{\kappa}_1^{*}\e^{\theta_2-\theta_1}.
\end{align*}

\renewcommand{\arraystretch}{1.5}
\begin{table}[!t]
	\centering
	\begin{tabular}{|c|c|}
		\hline
		\textbf{Soliton profiles} & \textbf{Existence condition} \\
		\hline
		$\mathrm{SHAD}$ & $d_{s,1}^{[1]}>0$  \\
		\hline
		$\mathrm{SVD}$ & $d_{s,1}^{[1]}<0$ \\
		\hline
		SHSV & $d_{s,1}^{[2]}d_{s,2}^{[2]}>0$ \\
		\hline
		DHAD & ~$d_{s,1}^{[3]}>0,d_{s,2}^{[3]}\ge 0$~\\
		\hline
		$\mathrm{DVD}$ & ~$d_{s,1}^{[3]}<0,d_{s,2}^{[3]}\le 0$~\\
		\hline
		$\mathrm{DHSV}$ & ~$d_{s,1}^{[3]}>0,d_{s,2}^{[3]}<0$~\\
		\hline
		$\mathrm{SHDV}$ & ~$d_{s,1}^{[3]}<0,d_{s,2}^{[3]}>0$~\\
		\hline
	\end{tabular}
	\caption{The correspondence between the seven basic intensity types and  existence condition. The corresponding full names of the soliton type abbreviations used in the table are as follows: SHAD: single hump anti-dark; SVD: single valley dark;  SHSV: single-hump-single-valley (dark-antidark complexes soliton); DHAD: double-hump antidark; DVD: double-valley dark; DHSV: double-hump-single-valley (anti-W-shaped soliton); SHDV: single-hump-double-valley (W-shaped soliton).}
	\label{tab:example}
\end{table}

\begin{figure}[!b]
	\centering		
	\includegraphics[width=1\columnwidth]{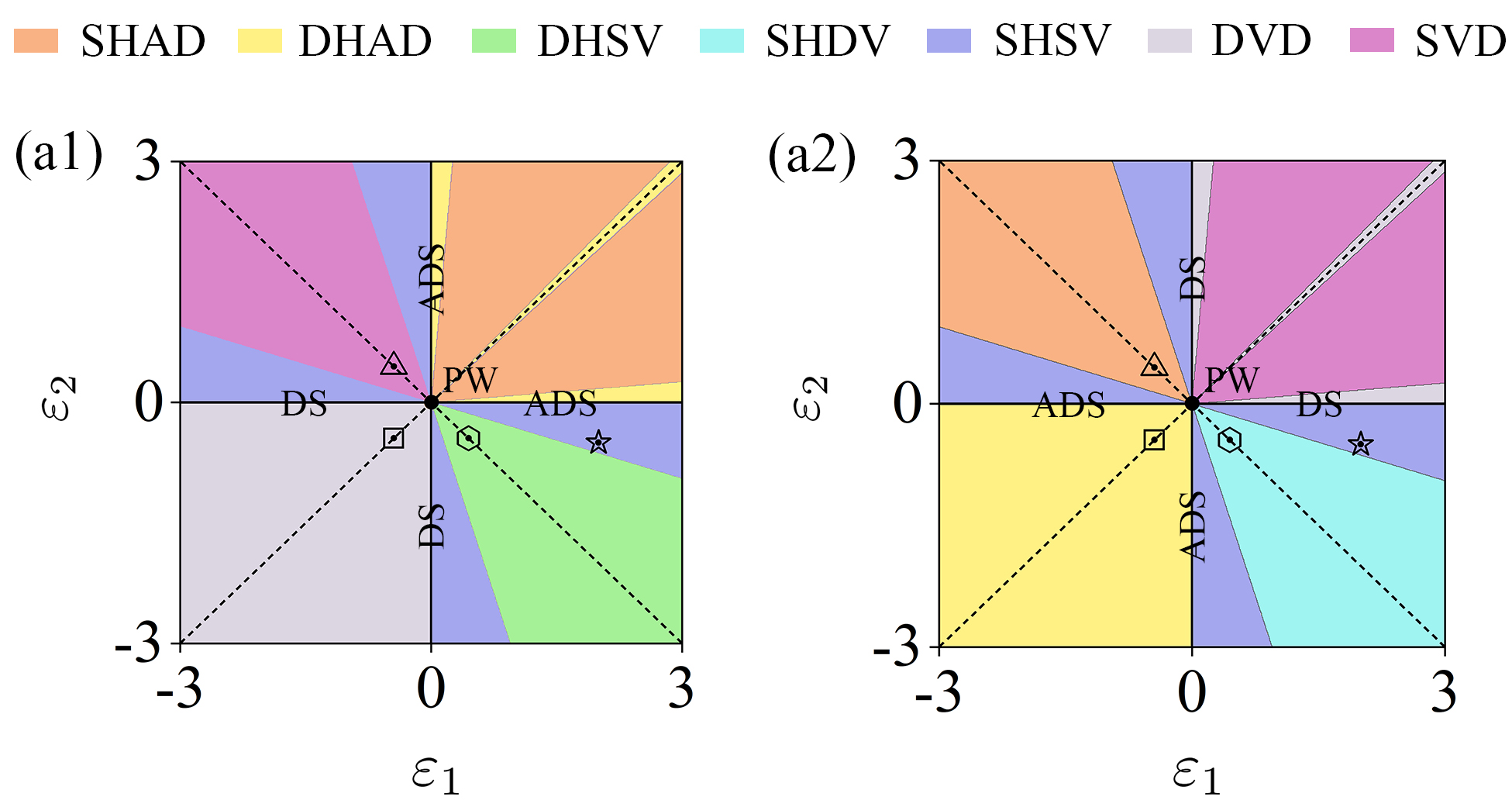}
	\caption{Intensity profiles of MHMVSs in the ($\varepsilon_1,\varepsilon_2$)-plane.  (a1) and (a2) correspond to the first and second components, respectively.  The black dash lines are symmetric intensity lines, which is given by $|\varepsilon_1|=|\varepsilon_2|$. The cube pattern point corresponds to Fig. \hyperref[fig1]{1(a)}; the star pattern corresponds to Fig. \hyperref[fig1]{1(b)}; the regular hexagonal pattern corresponds to Fig. \hyperref[fig1]{1(c)}; the triangular pattern corresponds to Fig. \hyperref[fig1]{1(d)}. The parameters are selected as follows:  $b_1=-b_2=0.5,z_1=0.5,z_2=1,a_1=\sqrt{5},a_2=1,\kappa_1=0.5+0.5\ii,\kappa_2=1+0.5\ii$. }
	\label{fig2}
\end{figure}

For convenience in analyzing the intensity structures, we consider the static case by setting $\xi=0$, $v=0$ without loss of generality. Some intensity structures of MHMVSs are shown in Fig. \ref{fig1}.  A convenient way to map out the possible profiles is to examine the number $m$ of extrema of $|u_s|^2$ as well as the sign of $d_{s,n}^{[m]} \equiv |u_s|_{\tau=\tau_{s,n}^{[m]}} - |a_s|$ at each extreme point $\tau_{s,n}^{[m]}$. The correspondence between the seven basic intensity types and $d_{s,n}^{[m]}$ for $m=1,2,3$ is shown in Table \ref{tab:example}. Besides, when $\Im(\kappa_1)\neq\Im(\kappa_2)$, the condition of the symmetric soliton intensity profile can be given as follows:
 \begin{align}\label{sym}
 	\ln\frac{\rho}{|\varepsilon_1\varepsilon_2|}=\frac{\Im(\kappa_2)+\Im(\kappa_1)}{\Im(\kappa_2)-\Im(\kappa_1)}\ln\frac{|\varepsilon_1|}{|\varepsilon_2|}.
 \end{align}
 For the case $\Im(\kappa_1)=\Im(\kappa_2),$ the soliton intensity profile will be symmetric when $|\varepsilon_1|=|\varepsilon_2|$. The axis of symmetry of the soliton intensity profile is  $\tau=[\Im(\kappa_1)+\Im(\kappa_2)]^{-1}\ln(\rho|\varepsilon_1\varepsilon_2|^{-1}).$
 As an example, the distribution of these seven intensity profiles in the $(\varepsilon_1,\varepsilon_2)$-plane for specific parameters is presented in Fig.~\ref{fig2}. The black dashed lines indicate where the intensity profile becomes symmetric.

\begin{figure}[!t]
	\centering		
	\includegraphics[width=85mm]{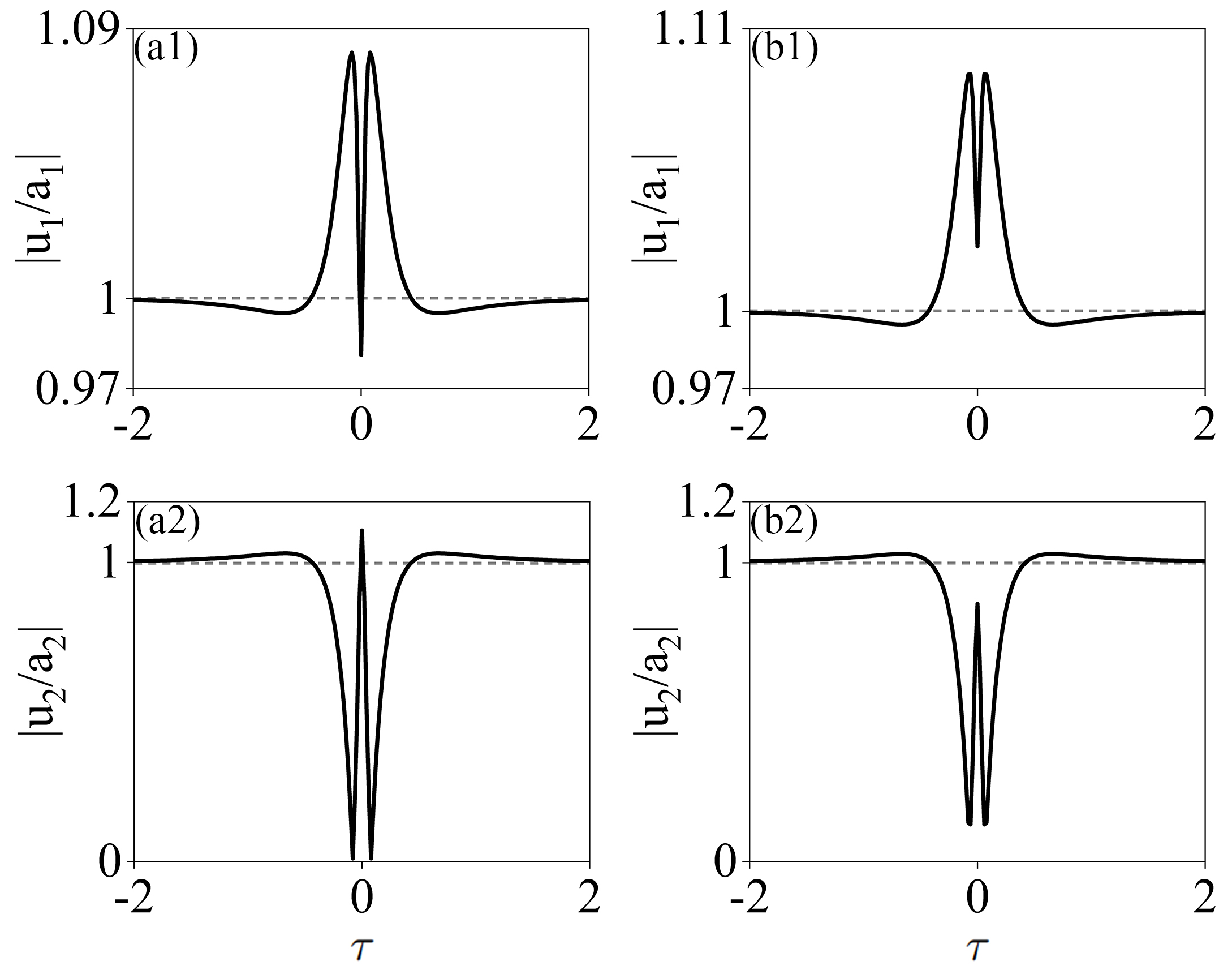}
	\caption{Some higher-order intensity structures of MHMVSs. (a1)-(a2) demonstrate symmetric DHTV-THDV soliton, (b1)-(b2) demonstrate symmetric DHDV-DHDV soliton.  The parameters are selected as follows: (a) $b_1=-b_2=2.7,z_1=6.25,z_2=\zeta^{-2}z_1\approx 5.6153,\zeta=1.055,a_1\approx 0.9452,a_2\approx 0.3907,$ $\kappa_1\approx 6.25+1.1443\ii,\kappa_2\approx 5.6153+2\ii,\varepsilon_1=-\varepsilon_2\approx-0.3321$;  (b) $b_1=-b_2=2.5,z_1=6.25,z_2=\zeta^{-2}z_1\approx 5.6153,\zeta=1.055,a_1\approx 0.9976,a_2\approx 0.4419,$ $\kappa_1\approx 6.25+1.0917\ii,\kappa_2\approx 5.6153+2\ii,\varepsilon_1=-\varepsilon_2\approx-0.3511$.}
	\label{fig3}
\end{figure}

The high-order solutions with more extreme points can appear in the second or fourth quadrant of the plane ($\varepsilon_1,\varepsilon_2$) for some given parameters.  However, these complex structures occur only in very narrow intervals of the parameter space; we do not attempt to chart them in a global phase diagram. Fig. \ref{fig3} shows some intensity structures with five extreme points that can emerge with some parameters, such as symmetric double-hump-triple-valley/triple-hump-double-valley (DHTV/THDV) structure in Fig. \hyperref[fig3]{\ref{fig3}(a)} and symmetric double-hump-double-valley/double-hump-double-valley (DHDV/DHDV) structure in Fig. \hyperref[fig3]{\ref{fig3}(b)}.  The intensity symmetry conditions are still given by Eq. (\ref{sym}). By varying the spatial position parameters $\varepsilon_1$ and $\varepsilon_2$ without changing the other parameters, one can control the structure of the soliton. This control method is similar to that used in Ref. \cite{Ren}, where varying the phase alters the intensity structure of the $P$ component in the nonlinear Schr\"{o}dinger-Maxwell-Bloch equation. In the CFL equations, both components will exhibit this transformation at the same time.

\subsection{Arbitrary Multihump-Multivalley Solitons}\label{sec2B}

\begin{figure*}[htp]
	\centering		
	\includegraphics[width=170mm]{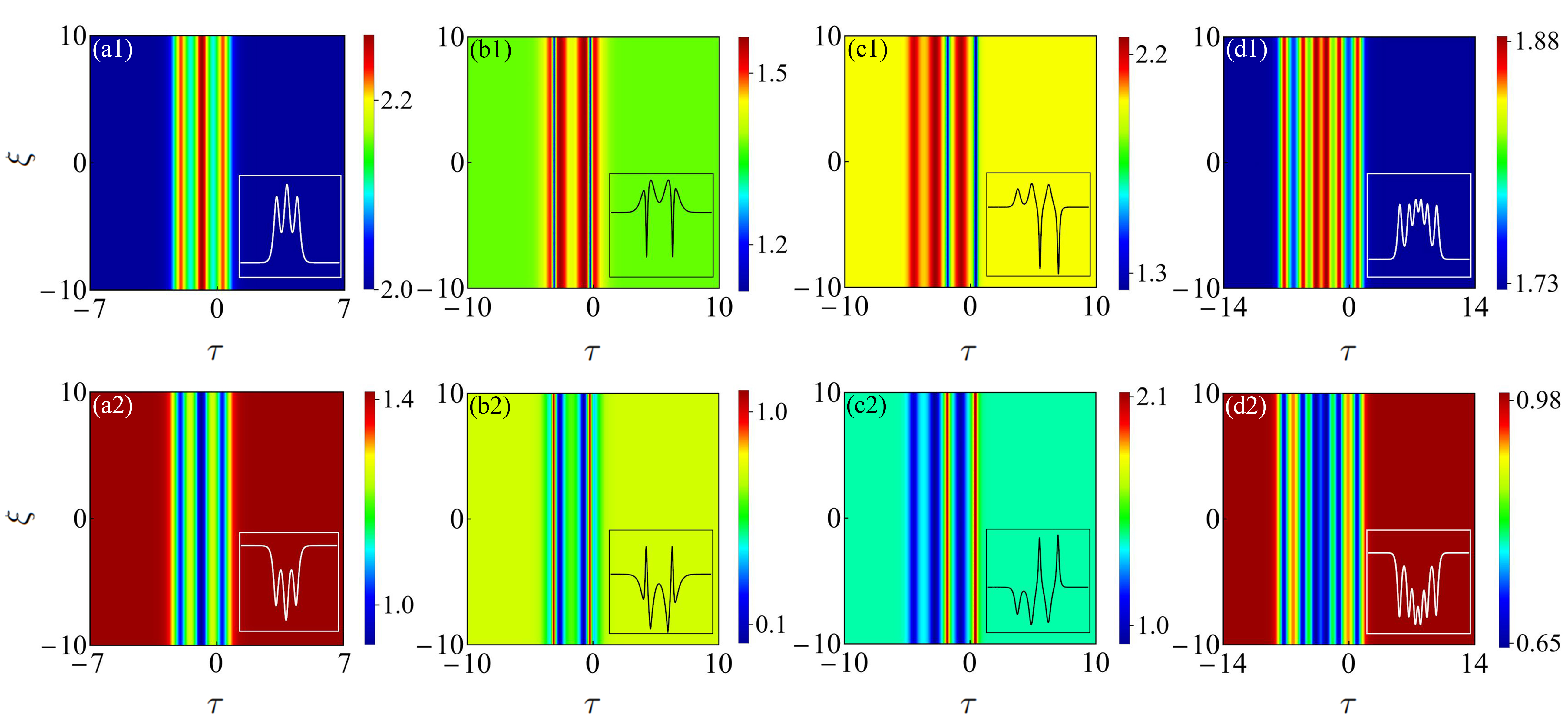}
	\caption{Some intensity distributions of static AMHMVSs. First row corresponds to the first component, and second row to the second component. (a) is obtained by three iterations with parameters  $\varepsilon_1=1,\varepsilon_2=0.5,\varepsilon_3=0.75$ and $a_1=2,a_2=\sqrt{2},b_1=-b_2=1,z_1=1,z_2=\frac{25}{16},z_3=\frac{16}{9},\kappa_1=1+2\ii,\kappa_2\approx 1.5625+2.4359\ii,\kappa_3\approx 1.7778+2.5507\ii$;  (b) is obtained by four iterations with parameters  $\varepsilon_1=1,\varepsilon_2=-0.39,\varepsilon_3=-3.2,\varepsilon_4=0.5$ and $a_1=\frac{5}{2\sqrt{2}},a_2=\frac{\sqrt{17}}{2},b_1=2b_2=2,z_1=1,z_2=\frac{25}{16},z_3=\frac{16}{9},z_4=\frac{36}{25},\kappa_1=1+1.5\ii,\kappa_2\approx 1.5625+1.9754\ii,\kappa_3\approx 1.7778+2.0898\ii,\kappa_4\approx 1.44+1.8964\ii$;  (c) is obtained by five iterations with parameters  $\varepsilon_1=4,\varepsilon_2=2,\varepsilon_3=-1,\varepsilon_4=2,\varepsilon_5=-0.1$ and $a_1=2,a_2=\sqrt{2},b_1=-b_2=1,z_1=1,z_2=\frac{25}{16},z_3=\frac{16}{9},z_4=\frac{36}{25},z_5=\frac{9}{4},\kappa_1=1+2\ii,\kappa_2\approx 1.5625+2.4359\ii,\kappa_3\approx 1.7778+2.5507\ii,\kappa_4\approx 1.44+2.3593\ii,\kappa_5\approx 2.25+2.7272\ii$; (d) is obtained by five iterations with parameters  $\varepsilon_1=3,\varepsilon_2=0.2,\varepsilon_3=1.3,\varepsilon_4=0.11,\varepsilon_5=26,\varepsilon_6=0.24$ and $a_1=\sqrt{\frac{149}{50}},a_2=\frac{7}{5\sqrt{2}},b_1=-b_2=1,z_1=1,z_2=\frac{25}{16},z_3=\frac{16}{9},z_4=\frac{36}{25},z_5=\frac{9}{4},z_6=\frac{49}{36},\kappa_1=1+1.4\ii,\kappa_2\approx 1.5625+1.6571\ii,\kappa_3\approx 1.7778+1.6969\ii,\kappa_4\approx 1.44+1.6214\ii,\kappa_5\approx 2.25+1.6875\ii,\kappa_6\approx 1.3611+1.5929\ii$.}
	\label{fig4}
\end{figure*}

The fundamental MHMVSs discussed above exhibit rich structures, yet the number of their humps and valleys is limited by the order of the DT. In deriving the velocity conditions, surprisingly, we have observed a notable feature encoded in Eq. (\ref{18}): for a fixed set of background parameters and a given spectral parameter $z_p$, there exists an infinite set of admissible spectral parameters $z_q$  satisfying the velocity condition. This infinite flexibility naturally raises the question of whether this freedom can be exploited to construct novel soliton structures.

Interestingly, we show that the answer is affirmative. Eqs. (\ref{14}) and (\ref{18}) show that, once $\Im(\kappa_p)$ and $z_p$ are fixed,  $z_q$ can be chosen freely within a certain interval, with $\kappa_q$ uniquely determined thereafter. The infinite admissible values of $z_q$ on this interval allow us to construct zero-velocity solitons with an arbitrary number of humps and valleys in the two-component coupled integrable system, by systematically exploiting this freedom and sequentially applying the DT. We refer to this class of solitons as AMHMVSs. The type of each soliton added by DT in the two components is determined by the sign of $b_s\varepsilon_h$, which can be concisely encoded in the matrix as
\setlength{\extrarowheight}{-2pt}
$$
\begin{pmatrix}
	b_1\varepsilon_1 & \cdots & b_1\varepsilon_N\\[-2pt]
	b_2\varepsilon_1 & \cdots & b_2\varepsilon_N 
\end{pmatrix}.
$$
Here $b_s\varepsilon_h>0$ and $b_s\varepsilon_h<0$ ($h=1,\cdots,N$) correspond to antidark and dark solitons, respectively, as established in Sec. \ref{sec2A}.
The $s$th row ($s=1,2$) of the matrix corresponds to the $s$th component, while the $h$th column corresponds to the soliton added at the $h$th DT. Besides, 
$a_1^2\!=\!\frac{[\Re(\kappa)+b_1]^2+[\Im(\kappa)]^2}{zb_1(b_1-b_2)}$ and $ a_2^2\!=\!\frac{[\Re(\kappa)+b_2]^2+[\Im(\kappa)]^2}{zb_2(b_2-b_1)}$ imply $zb_1\Delta b>0,~-zb_2\Delta b>0$, showing that $b_1b_2<0.$ AMHMVSs allow the structure of $N$ humps or $N$ valleys, as shown in Fig. \hyperref[fig4]{\ref{fig4}(a)} and Fig. \hyperref[fig4]{\ref{fig4}(d)}, as well as the interlocking structure of multihumps and multivalleys, as shown in Fig. \hyperref[fig4]{\ref{fig4}(b)} and Fig. \hyperref[fig4]{\ref{fig4}(c)}. 

In previous DT studies of coupled nonlinear Schrödinger systems \cite{DV-ling1,Qin1,Qin2,NS-Stalin2,NS-Stalin1,NS-Lin,zhao3,DV-ling,Qin}, solitons with multiple humps or valleys were constructed via velocity resonance, yet the number of extrema was inherently limited by the order of the DT and the available spectral parameters. The emergence of AMHMVSs in the present work is therefore not merely an extension to higher-order DT iterations. Rather, it originates from a fundamentally new ingredient: the infinite admissible choices of the spectral parameter. This infinite flexibility enables the superposition of an arbitrary number of DS and ADS constituents within the same component, in stark contrast to nondegenerate solitons, which arise from incoherent superposition across different components \cite{Qin1}. As constructed via the DT, AMHMVSs possess explicit analytic expressions and admit infinitely many distinct structural types. The free parameter $\varepsilon_h$ directly controls the position of each hump or valley, providing a flexible means to tune the overall soliton profile.

\section{Phase Properties of Fundamental Multihump-Multivalley Soliton}\label{sec3}

Phase plays a significant role in the physical properties of nonlinear waves. Recent studies within the topological vector potential framework have revealed that the intensity zeros of dark solitons \cite{Zhao3} and rogue waves \cite{Zhao5} give rise to virtual monopoles with quantized flux $\pm\pi$. In the Hirota and Sasa-Satsuma systems, it was demonstrated that rational W-shaped solitons can possess completely different phase structures in the two systems \cite{Zhao2}, whereas their intensity profiles are nearly identical, highlighting the sensitivity of topological phases to the underlying model. Besides, the dark solitons with the same velocity, which possess virtual monopoles with quantized flux $\pm\pi$ corresponding to intensity zeros, can admit two distinct phase shifts in the Hirota equation \cite{Qin3}. By tracing the intensity zeros in real space using Dirac's theory, magnetic monopoles with charges $\pm 1$ have been identified in the Akhmediev breather \cite{Yu}. Recent investigations have further revealed that the topological charge of solitonic monopoles depends critically on the nature of the nonlinearity: in the Chen-Lee-Liu system, where both cubic nonlinearity and self-steepening are present, solitons carry $\pm\frac{1}{2}$ monopole charges \cite{Wu}, whereas in the derivative nonlinear Schr\"{o}dinger system, where the self-steepening effect acts alone, the charge can be elevated to $\pm\frac{3}{2}$ for simple intensity poles and further to $\pm\frac{5}{2}$ for third-order ones \cite{arXiv1}. This naturally raises the question of whether MHMVSs, driven by the interplay of self-steepening and spatiotemporal coupling, host distinct and previously unexplored phase structures. 

\begin{figure}[t]
	\centering		
	\includegraphics[width=0.75\columnwidth]{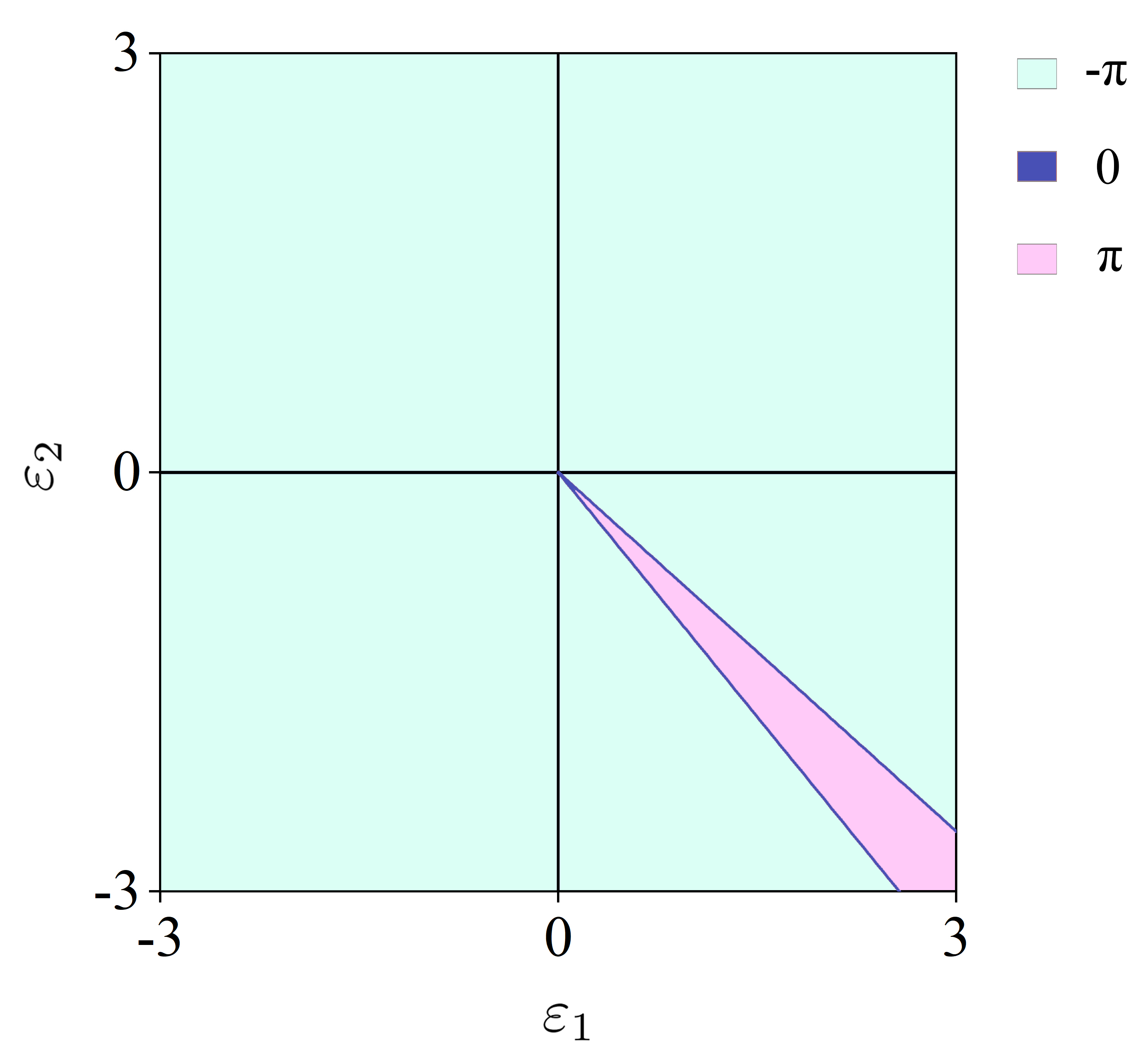}
	\caption{Phase-shift difference between the two components for $(\varepsilon_1,\varepsilon_2)\in(\mathbb{R}\backslash\{0\})^2$ at $\xi=0$.  The parameters are same as Fig. \ref{fig2}. }
	\label{fig5}
\end{figure}

\begin{figure*}[htp]
	\centering		
	\includegraphics[width=1.9\columnwidth]{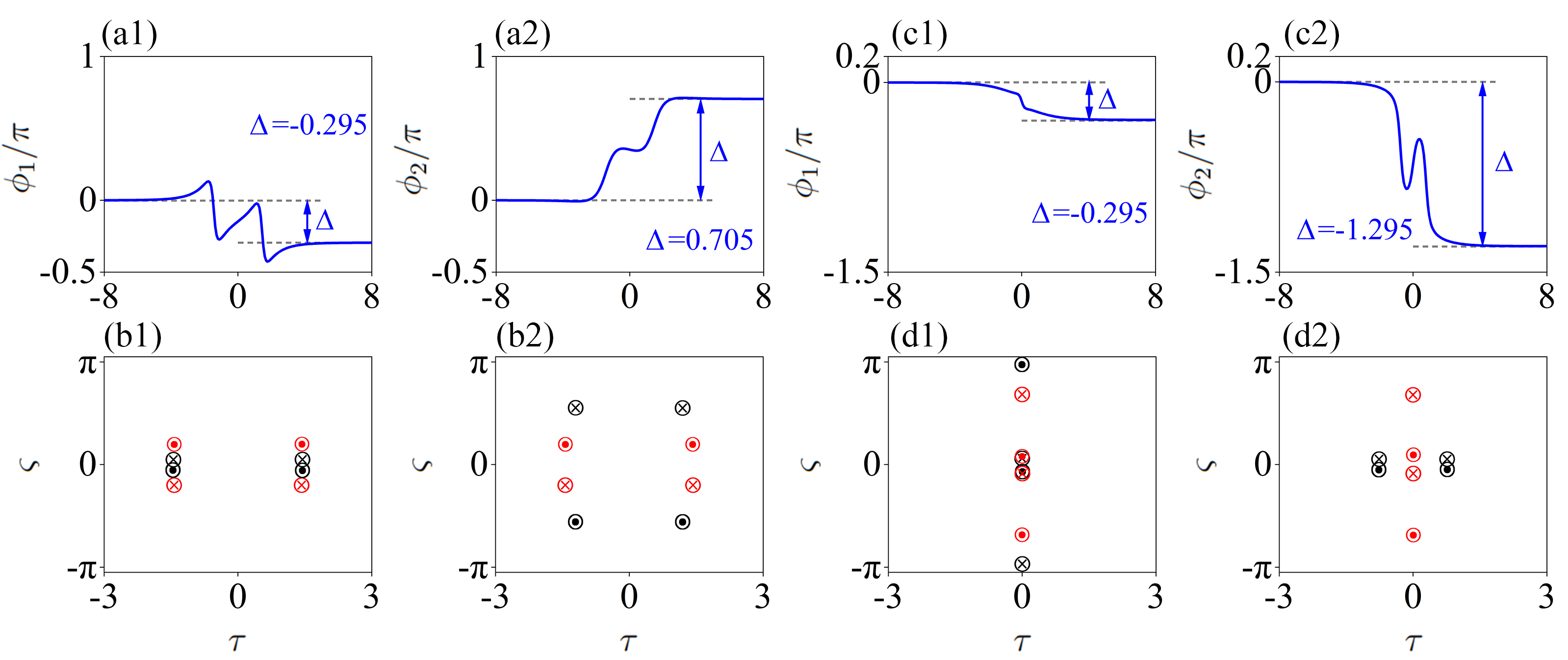}
	\caption{Phase properties of MHMVSs. Panels (a1)-(a2) and (b1)-(b2)  show the phase distribution function associated with Fig. \hyperref[fig1]{1(a1)-(a2)} and its corresponding topological vector potential, respectively. Panels (c1)-(c2) and (d1)-(d2)  show the phase distribution function associated with Fig. \hyperref[fig1]{1(c1)-(c2)} and its corresponding topological vector potential, respectively. $\Delta$ denote the value of corresponding phase shift $\Delta \phi$. The monopoles with positive and negative charges are denoted by $\odot$ and $\otimes$ and the value is $\pm\frac{1}{2}$, respectively.  The red symbols in (b1)-(b2) and (d1)-(d2) correspond to the intensity poles, while the black symbols correspond to intensity zeros.}
	\label{fig6}
\end{figure*}

To this end, we start from the phase function of the soliton \eqref{b_mhmv_sol}, which is  defined as $\phi_s=\arg\frac{\Lambda^{[s]}}{\Gamma}$ \cite{Wu,Zhao2,Zhao3,Yu,Zhao5,Zhao6,arXiv1,Qin3}. It can be rewritten as $\phi_s=\arctan\frac{\Im(\pmb{\mathrm{\Omega}}_s)}{\Re(\pmb{\mathrm{\Omega}}_s)}+n\pi$ as $\Re(\pmb{\mathrm{\Omega}}_s)\neq0$, where $\pmb{\mathrm{\Omega}}_s=\Lambda^{[s]}\Gamma^*$. Here, $n$ is a piecewise constant function of $\tau$ that jumps when $\Re(\pmb{\mathrm{\Omega}}_s)=0$ to ensure the continuous extension of the phase across the singularity. The phase gradient flow can be given as
\begin{align}\label{pgf}
	\phi_{s,\tau}\!=\!\frac{\Re(\pmb{\mathrm{\Omega}}_s)\partial_\tau\Im(\pmb{\mathrm{\Omega}}_s)\!-\!\Im(\pmb{\mathrm{\Omega}}_s)\partial_\tau\Re(\pmb{\mathrm{\Omega}}_s)}{|\pmb{\mathrm{\Omega}}_s|^2}
\end{align}
as $\Re(\pmb{\mathrm{\Omega}}_s)\neq 0$. If $\Re(\pmb{\mathrm{\Omega}}_s)=0$ at a certain point $\tau=\tau_0$, we continuously extend $\phi_{s,\tau}$ at this point by taking the limit. In addition, the phase function can be reformulated into a continuous form $\phi_s(\tau)=\int_{-\infty}^\tau\phi_{s,\tau}\dd \tau$, and the phase shift $\Delta\phi_s=-\lim\limits_{\tau\rightarrow+\infty}\phi_s(\tau)$. Direct calculation shows that the asymptotic phase limits $\lim\limits_{\tau\rightarrow\pm\infty}\phi_{s}(\tau)$ are independent of the parameters $\varepsilon_1$ and $\varepsilon_2$. However, as demonstrated in Fig. \ref{fig2}, these parameters strongly affect the intermediate profile of the soliton. Since solitons with different intensity structures exhibit distinct phase variation curves, a more detailed analysis of the phase properties is required. To this end, we define $\varepsilon_2=\varpi\varepsilon_1,x=\e^{\tau}\varepsilon_1$, then the phase gradient flow Eq. \eqref{pgf} can be expressed as a form of  $\phi_{s,\tau}\equiv\frac{xf_s(x,\varpi)}{\varepsilon_1g_s(x,\varpi)},$ where $f_s$ and $g_s$  are finite sums of the form $\sum_{n}C_{s,n}(\varpi)x^{\beta_n}$ with $\beta_n\ge 0$. Besides, we need to restrict $(\varpi,\varepsilon_1)\in(\mathbb{R}\backslash\{0\})^2$ to prevent the solution from degenerating into a single soliton. Moreover, the phase shift of each component can be calculated by
\begin{align}\label{dp}
	\Delta\phi_s=\int_{-\infty}^\infty\phi_{s,\tau}\dd\tau=\int_{0}^{\mathrm{sgn}(\varepsilon_1)\infty}\frac{f_s(x,\varpi)}{g_s(x,\varpi)}\dd x.
\end{align}
To simplify our analysis, we take the corresponding parameter case in Fig. \ref{fig2} as an example. The exact expressions of $f_s$ and $g_s$ are given in Appendix \ref{App2}. To determine whether $f_s$ and $g_s$ share common factors, we compute their resultant $R(f_s,g_s)$.
For $s=1$, $R(f_1(x),g_1(x))\neq 0$ for all $\varpi\in\mathbb{R}\backslash\{0\},$ while for $s=2$, $R(f_2(x),g_2(x))=0$ as $\varpi=\varpi_\pm=\frac{-16\pm\sqrt{6}}{5\sqrt{10}}$, and $\mathrm{gcd}(f_2(x,\varpi_\pm),g_2(x,\varpi_\pm))=(x-x_\pm)^2$, where $x_\pm=\frac{1}{5}(2\sqrt{2}\pm\sqrt{3}).$ 
Since $x_\pm>0,\varpi_\pm<0$, singularities occur only for $\varepsilon_1>0$ and $\varepsilon_2<0$, i.e., in the fourth quadrant of the $(\varepsilon_1,\varepsilon_2)$-plane. Figure \ref{fig5} displays the distribution of the phase shift difference $\Delta\phi_1-\Delta\phi_2$ in $(\varepsilon_1,\varepsilon_2)$-plane. In the fourth quadrant, two distinct rays are observed where the phase shift difference vanishes; their slopes are precisely $\varpi_+$ and $\varpi_-$. Between these two rays, the phase shift difference jumps to $\pi$, while outside this interval it takes the value $-\pi$. This piecewise behavior originates from the singularity removal caused by the limiting cancellation. Quantitatively, the phase-shift difference between two components is given by
\begin{equation}\label{Dp_cfl}
	\Delta\phi_1-\Delta\phi_2=\!\left\{
	\begin{aligned}
		&\pi,&&\!\varpi\in(\varpi_-,\varpi_+), \varepsilon_1>0,\varepsilon_2<0,&\\
		&0,&&\!\varpi=\varpi_\pm, \varepsilon_1>0,\varepsilon_2<0,&\\
		&-\pi,&&\!\mathrm{other~points}.&
	\end{aligned}
	\right.
\end{equation} 
As representative examples, Figs. \hyperref[fig6]{\ref{fig6}(a1)-(a2)} ($\varepsilon_1=\varepsilon_2=-\frac{1}{\sqrt{5}}$) and Figs. \hyperref[fig6]{\ref{fig6}(c1)-(c2)} ($\varepsilon_1=-\varepsilon_2=\frac{1}{\sqrt{5}}$) show the phase distributions of the two components, which exhibit markedly different trends in these two cases.

To understand the underlying mechanism of these phase evolutions, we further analyze the phase properties of MHMVSs through the topological vector potential theory \cite{Wu,Zhao2,Zhao3,Yu,Zhao5,Zhao6,arXiv1,Qin3}. Consider complex extension $\tau\mapsto z= \tau+\ii\varsigma$. The topological vector potential can be given by $\mathbf{A}_s=\Re\left[\frac{\partial\phi_s(z)}{\partial \tau}\mathbf{e}_\tau+\frac{\partial\phi_s(z)}{\partial \varsigma}\mathbf{e}_\varsigma\right]=\Re[{\partial_z\phi_s(z)}]\mathbf{e}_\tau-\Im[\partial_z\phi_s(z)]\mathbf{e}_\varsigma$. 
For the case where the magnetic monopole distribution has a period $T$ along the imaginary axis (in above case, $T=2\pi$), with the properties of meromorphic functions, the topological vector potential of $\partial_z\phi_s$ can be expressed as:
\begin{align}\label{cfl_tvp}
	\mathbf{A}_s=\sum_{n=-\infty}^{\infty}&\sum_{j=1}^{r_s}\bigg[-\frac{1}{2} \frac{(\tau-\tau_{s,j}^-)\mathbf{e}_\varsigma-[\varsigma-(\varsigma_{s,j}^-+nT)]\mathbf{e}_\tau}{(\tau-\tau_{s,j}^-)^2+[\varsigma-(\varsigma_{s,j}^-+nT)]^2}\nonumber\\
	&+\frac{1}{2} \frac{(\tau-\tau_{s,j}^+)\mathbf{e}_\varsigma-[\varsigma-(\varsigma_{s,j}^++nT)]\mathbf{e}_\tau}{(\tau-\tau_{s,j}^+)^2+[\varsigma-(\varsigma_{s,j}^++nT)]^2}\bigg],
\end{align}
where $r_s$ is the number of singularities of $\partial_z\phi_s(z)$ and points $(\tau_{s,j}^\pm, \varsigma_{s,j}^\pm)$ are their coordinates in the range $(\tau,\varsigma)\in\mathbb{R}\times \left[-\frac{T}{2}, \frac{T}{2}\right]$, and sign $\pm$ corresponds to the sign of the charges. As shown in Figs. \hyperref[fig6]{\ref{fig6}(b1)-(b2)} and Figs. \hyperref[fig6]{\ref{fig6}(d1)-(d2)}, within the magnetic field corresponding to a period interval $[-\pi,\pi]$, the magnetic monopoles (black and red symbols) all have charge $\pm\frac{1}{2}$. The coordinates of these magnetic monopoles are determined by the singularities of the topological vector potential (i.e., the singularities of the phase gradient flow). In general, the singularities of the phase gradient flow are associated with the poles and zeros of the intensity function on the complex plane. In fact, the singularities of the phase gradient flow originate from the roots of $|\pmb{\Omega}_s|^2=0$: those from 
$|\Gamma|^2=0$ correspond to intensity poles, while those from $|\Lambda^{[s]}|^2=0$ correspond to intensity zeros. When  a limiting cancellation occurs (i.e., the right-hand side takes an indeterminate form $0/0$ or $\infty/\infty$), some singularities are removed; otherwise, all are retained. In either case, the correspondence between singularities and intensity zeros/poles remains unchanged.

In the periodic case, both $\Gamma$ and $\Lambda^{[s]}$ can be reduced to polynomials in $\e^{\tau}$. Within a single period of the magnetic monopole distribution, the number of singularities is directly determined by the polynomial degree. In our example, for $\phi_{1,\tau}$, there is no limiting cancellation in Eq. \eqref{pgf}. Within one period,  $|\Gamma|^2=0$ determines four roots corresponding to intensity poles, while $|\Lambda^{[s]}|^2=0$ determines four roots corresponding to intensity zeros. As shown in Figs. \hyperref[fig6]{\ref{fig6}(b1)-(b2)} and Figs. \hyperref[fig6]{\ref{fig6}(d1)-(d2)}, the magnetic monopoles marked in red correspond to poles, while those marked in black correspond to zeros, and within one period, there are four of each. For $\phi_{2,\tau}$, the limiting cancellation results in the removal of two roots of $|\pmb{\mathrm{\Omega}}_s|^2=0$ from the singularity set as $\varpi=\varpi_\pm$. Since these two roots are a conjugate pair that coalesce on the real axis as parameters vary, they must originate from the same equation, either $|\Gamma|^2=0$ or $|\Lambda^{[2]}|^2=0$. Here these two roots come from the equation $|\Lambda^{[2]}|^2=0$, corresponding to the coalescence of two zeros of the intensity function, thereby generating a zero of order two in the intensity function. Furthermore, from solution \eqref{b_mhmv_sol}, the magnetic monopoles associated with the intensity poles in the two components are identical, both being determined by $|\Gamma|^2=0$. 

In fact, for the intensity function determined by solution \eqref{solution}, $\det H^{[s]}$ and $\det H$ share the same algebraic structure about $\e^{\Im(\kappa_i)(\tau-v_i\xi)+\Im(\kappa_j)(\tau-v_j\xi)}$. Therefore, as long as the parameters are chosen to ensure a finite period in the magnetic monopole distribution (which is achieved when $\frac{\Im(\kappa_i)}{\Im(\kappa_j)}\in\mathbb{Q}$ for any $\kappa_i,\kappa_j$), then this periodicity ensures that the expressions for the numerator and denominator of the intensity function can be transformed into the same polynomial structure, so the numbers of zeros and poles must be equal. That is, exactly half of the magnetic monopoles correspond to the zeros of the intensity, and the other half correspond to the poles of the intensity. These zeros and poles correspond to potential monopole contributions; however, whether the monopole field is truly equally partitioned depends on whether any singularities are removed by cancellation in the phase gradient flow. Specifically, let $ P_s = \Re(\det H^{[s]}\det H^*) $ and $ Q_s = \Im(\det H^{[s]}\det H^*)$ denote the polynomials associated with the real and imaginary parts of the phase function. If $ \gcd(P_s,Q_s)=1 $, all zeros and poles contribute to the monopole field and the equal partition holds; if a nontrivial common factor exists, the corresponding singularities are removed. Moreover, if the numbers of canceled roots from $ \det H^{[s]} $ and $\det H$ are the same, the equal partition is preserved after reduction; if they differ, it is broken. That is, exactly half of the magnetic monopoles correspond to intensity zeros and the other half to intensity poles, provided that the above no-cancellation or symmetric-cancellation condition is satisfied. Thus, the same equal-partition property is expected to hold for AMHMVSs under the same algebraic conditions.

Next, we analyze the cause of the phase shift from the perspective of magnetic monopole motion. In fact, $\Delta\phi_1$ remains invariant over the entire $(\varepsilon_1,\varepsilon_2)$ plane ($\varepsilon_1,\varepsilon_2\neq 0$), and the result of Eq. \eqref{Dp_cfl} is completely determined by $\Delta\phi_2$. In $\Delta\phi_2$, since the functions $f_2$ and $g_2$ share a common factor under specific parameters, their cancellation leads to a sudden change in the phase shift. In terms of the topological vector potential, this cancellation corresponds to the merging of monopoles on the real axis. Such merging acts on the phase and affects the phase shift. The phase shift in the fourth quadrant of Fig. \ref{fig5} reflects the following motion logic of the magnetic monopoles: starting from the initial state $\Delta\phi_1-\Delta\phi_2=-\pi$ for $\varpi<\varpi_-$, when a magnetic monopole moves onto the real axis (i.e., $\varpi=\varpi_-$), the factor corresponding to $x_-$ is canceled, reducing $\Delta\phi_2$ by $\pi$; then, as $\varpi$ evolves into the interval $(\varpi_-,\varpi_+)$, the previously disappeared monopole pair crosses the real axis and reappears, moving to the opposite half-planes from their original ones, causing the phase shift to further decrease by $\pi$; until $\varpi=\varpi_+$, the factor corresponding to $x_+$ is canceled, increasing the phase shift by $\pi$; finally, for $\varpi>\varpi_+$, similar to before, the monopole pair each crosses the real axis, further increasing the phase shift by $\pi$, thereby returning the phase shift to its initial value.

This monopole motion picture, while physically transparent, can be made rigorous by integrating the topological vector potential \eqref{cfl_tvp} along the real axis.
Under the Cauchy principal value prescription, this integration yields an expression for the phase shift in terms of the monopole positions and charges \cite{arXiv1}
\begin{align}\label{dp_s2}
		\Delta\phi_s
		=-\frac{2\pi}{T}\sum_{j=1}^{r_s}\bigg[&-\frac{1}{2}\left(\mathrm{sgn}(\varsigma_{s,j}^-)\frac{T}{2}-\varsigma_{s,j}^-\right)\nonumber\\
		&+\frac{1}{2}\left(\mathrm{sgn}(\varsigma_{s,j}^+)\frac{T}{2}-\varsigma_{s,j}^+\right)\bigg],
\end{align} 
where $\varsigma_{s,j}^+=-\varsigma_{s,j}^-.$ This expression is evidently piecewise, with discontinuities that arise exactly from the merging of magnetic monopoles on the real axis. Analyzing the limits on both sides of the discontinuities reveals that the phase shift undergoes two distinct jumps:  the first as the monopole pair arrives and merges on the real axis, and the second as they re-emerge on the opposite side after crossing. For the present case where the monopole charges are $\pm\frac{1}{2}$, each jump gives rise to an absolute phase shift of $\pi$. Indeed, for a monopole with charge $\pm\sigma$, the same process would yield an absolute phase shift of $2\sigma\pi$.

Therefore, we have systematically analyzed the phase properties of fundamental MHMVSs within the topological vector potential framework. This equal-partition feature of virtual magnetic monopoles stands in sharp contrast to previous studies \cite{Wu,Zhao2,Zhao3,Yu,Zhao5,Zhao6,arXiv1,Qin3}, and reflects the unique topological structure by the coherent superposition of dark and antidark constituents within the same component.

\section{Numerical evolution of Multihump-Multivalley solitons}\label{sec4}

\begin{figure*}[t]
	\centering		
	\includegraphics[width=160mm]{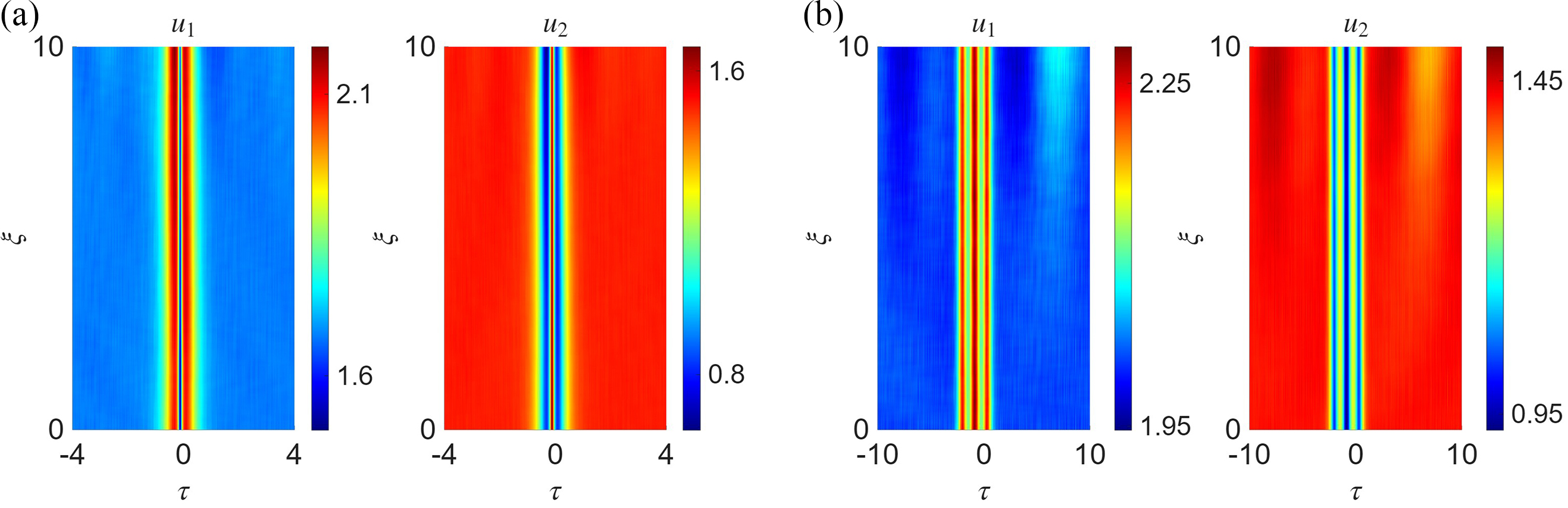}
	\caption{Numerical evolution diagram of MHMVSs with 1\% white noise. (a) Numerical evolution of anti-W-shaped/W-shaped soliton soliton. Parameters are given by: $\varepsilon_1=-4,\varepsilon_2\approx 0.9682, a_1\approx 1.6607,a_2\approx 1.5026, b_1=-b_2=4,z_1=1,z_2=0.25,\kappa_1\approx 1+7.9530\ii,\kappa_2=0.25+2\ii $; (b) Numerical evolution corresponding to Figs. \hyperref[fig4]{\ref{fig4}(a1)-(a2)}.}\label{fig7}
\end{figure*}

The analytical solutions derived in the previous sections are constructed under ideal conditions, assuming a perfect plane-wave background and an infinite spatial domain. However, for these theoretical findings to be relevant to experimental realizations in birefringent fibers, it is necessary to test their dynamical stability against perturbations that are unavoidable in real-world settings. An analytical solution that is unstable would be of little practical significance.

The primary purpose of this section is to numerically validate the robustness of MHMVSs and AMHMVSs. We take $\xi=0$ as the initial state and consider evolving system \eqref{1} over a given interval $I_0=[I_-,I_+]$. For the D-AD soliton of the coupled system \eqref{1}, it is difficult to find a common period, which prevents the Fourier spectral method from being directly applied to both components simultaneously. To address this issue, we discuss wavefunctions $u_s(\tau)$ with a phase difference  $\arg(u_s(I_-))-\arg(u_s(I_+))$. Consider the transformation
\begin{align}\label{tr_p}
	u_s\mapsto u_s\mathrm{e}^{\mathrm{i} k_s \tau}.
\end{align}	
By choosing appropriate $k_s$ to correct the phase difference of the original wavefunction $u_s$ at the two ends of the interval, periodic boundary conditions can be achieved. To balance this phase difference, the factor $\mathrm{e}^{\mathrm{i} k_s \tau}$ needs to satisfy
$\arg\left[u_s(I_-)\mathrm{e}^{\mathrm{i} k_s I_-}\right]=\arg\left[u_s(I_+)\mathrm{e}^{\mathrm{i} k_s I_+}\right],$
leading to $\arg(u_s(I_-))+k_s I_-=\arg(u_s(I_+))+k_s I_++2\pi\mathbb{Z}.$
Taking the integer as zero yields
$$k_s=\frac{\arg(u_s(I_-))-\arg(u_s(I_+))}{I_+-I_-},$$
thus the introduced correction phase is evenly distributed over the interval to achieve periodic boundaries.

Under the transformation \eqref{tr_p}, system \eqref{1} can be transformed into
\begin{subequations}\label{CFL2}
	\begin{align}
		D_{1,\tau}u_{1,\xi}=&- u_1-\ii\left(|u_1|^2+\frac{1}{2}\sigma|u_2|^2\right)D_{1,\tau}u_{1}&\nonumber\\
		&-\frac{\ii}{2}\sigma u_1u_2^*D_{2,\tau}u_{2},\\
		D_{2,\tau}u_{2,\xi}=&- u_2-\ii\left(\frac{1}{2}|u_1|^2+\sigma|u_2|^2\right)D_{2,\tau}u_{2}&\nonumber\\
		&-\frac{\ii}{2} u_2u_1^*D_{1,\tau}u_{1},
	\end{align}
\end{subequations}
where $D_{s,\tau}=\partial_\tau-\ii k_s.$ Next, we use the Fourier spectral method to evolve system \eqref{CFL2} in $\xi$. Under periodic boundary conditions, by means of the Fourier transform we have $D_{s,\tau} u_{s}(\tau,\xi)=\mathscr{F}^{-1}[\ii (k-k_s)\mathscr{F}(u_s)(k,\xi)](\tau).$  
Due to the presence of mixed derivative terms $D_{\tau}u_{s,\xi}$, system \eqref{CFL2} cannot directly evolve the wavefunctions $u_s$, but numerical methods such as the Runge-Kutta method can be used to evolve and obtain $D_{s,\tau} u_{s}(\tau)$ at the next step, and then integrate them to obtain the wavefunctions $u_s$ required for further evolution.

 We now examine the evolution of the soliton solutions under $1\%$ white noise. The addition of noise inevitably breaks the periodicity at the boundaries; nevertheless, the phase matching performed earlier ensures that the boundary conditions remain approximately periodic under $1\%$ white noise. Even so, errors arising from the non-periodic boundary components gradually propagate inward over time. Since these errors take a finite time to reach the central region, the evolution remains reliable within this time window. It should be noted that the presence of noise also makes the integration constants used to recover $u_s$ difficult to determine; we therefore adopt $\mathscr{F}(u(\tau,\xi_{N}))(0)$  as the integration constant at each iteration, as a working approximation. As an example, numerical simulation results of anti-W-shaped/W-shaped soliton and three-hump ADS/three-valley DS are shown in Fig. \ref{fig7}. The evolution interval $I_0$ is chosen sufficiently large so that boundary effects do not affect the central region within the simulation time; the plots are restricted to the central subinterval, and the evolution time is 10 units in all cases.  The results demonstrate that the soliton structures are well preserved under $1\%$ white noise.

\section{Conclusion}\label{sec5}

We analytically study MHMVSs on a plane-wave background for the CFL equations, an integrable model describing femtosecond pulse propagation in birefringent fibers with self-steepening and spatiotemporal coupling. Using the DT method, we systematically derive exact analytical MHMVS solutions, covering fundamental, higher-order, and static families with arbitrary hump/valley numbers. We classify seven fundamental intensity profiles and map out their existence domains in the parameter plane. In contrast to nondegenerate solitons formed by incoherent cross-component superposition, the MHMVSs reported here arise from coherent intra-component superposition of dark and antidark soliton constituents. Notably, we demonstrate for the first time that the static propagation regime supports MHMVS with an arbitrary number of humps and valleys. This result, previously unreported in two-component coupled integrable systems, significantly extends the class of exact localized wave solutions for vector nonlinear models. We further uncover a unique equal-partition topological phase property: intensity zeros and poles contribute equally to the virtual magnetic monopole field over one spatial period, a topological feature unprecedented among all previously studied nonlinear wave systems. Numerical simulations demonstrate that both fundamental and arbitrary MHMVS solutions remain structurally stable against weak white noise, lending support to their possible experimental realization.

It should be mentioned that MHMVSs are collision-forbidden with solitons of distinct velocities in this system. This follows from the amplitude constraint: for any third spectral parameter with the velocity different from the resonant pair, the system admits no positive real solutions for the squared imaginary parts of the associated spectral parameters. The same conclusion holds for configurations involving more than three spectral parameters. Extending the present two-component construction to the general $N$-component case would be of particular interest. Beyond the present scope, it would be natural to examine whether the equal-partition property persists for arbitrary 
$N$, how additional components modify the pole-zero structure, and the individual influence of self-steepening and spatio-temporal coupling on the topological phase.

\section{Acknowledgments}

Qin is grateful to Li-Chen Zhao, and Tao Jiang for their helpful discussions. This work is supported by the National Natural Science Foundation of China (Grant No. 12405004), the Natural Science Foundation of Xinjiang Uygur Autonomous Region Project (Grant No. 2024D01C232), the Scientific Research Projects Funded by the Basic Research Business Expenses of Autonomous Region Universities (Grant No. XJEDU2024P011), and the ``Tianchi Talent" Introduction Plan in Xinjiang Uygur Autonomous Region.

\appendix
\section{The derivation of general $N$-soliton solitons on the plane wave background}\label{App1}
In this section,  we present the procedure for deriving the $N$-soliton solutions of system \eqref{1} on the plane-wave background $u_s=a_s\e^{\ii \omega_s }$.
The Lax pair of system \eqref{1} is given by \cite{su,Ling1}
\begin{subequations}\label{Laxpair1}
	\begin{align}
		\Phi_\tau&=U\Phi,~~U=\ii\lambda^{-2}\sigma_3+\lambda^{-1}Q_\tau,\\
		\Phi_\xi&=V\Phi,~~V=\ii\left(\frac{1}{4}\lambda^2\sigma_3+\frac{1}{2}\sigma_3(Q^2-\lambda Q)\right),
	\end{align}
\end{subequations}
where $\lambda$ is the spectral parameter,  and matrices
$$\sigma_3=\begin{pmatrix}
	1 & 0 &  0\\
	0 & -1 & 0 \\
	0 & 0 & -1
\end{pmatrix},~Q=\begin{pmatrix}
	0 & u_1^* &  u_2^*\\
	u_1 & 0 & 0 \\
	u_2 & 0 & 0
\end{pmatrix}.$$
Setting $\Phi=D\Phi_0$ with $D=\diag(1,\e^{\ii\omega_1},\e^{\ii\omega_2})$, system (\ref{Laxpair1}) reduces to
\begin{subequations}\label{Laxpair2}
	\begin{align}
		\Phi_{0,\tau}&=\ii U_0\Phi_0,\\
		\Phi_{0,\xi}&=\ii V_0\Phi_0,
	\end{align}
\end{subequations}
where
\begin{align*}
	&U_0=\lambda^{-2}
	\begin{pmatrix}
		1 & -a_1b_1\lambda & - a_2b_2\lambda\\
		a_1b_1\lambda & -1-b_1\lambda^2 & 0 \\
		a_2b_2\lambda & 0 & -1-b_2\lambda^2
	\end{pmatrix},
\end{align*}	
\begin{align*}
	&V_0=\frac{\lambda^2}{2b_1b_2}U_0^2+\frac{(b_1+b_2)}{2b_1b_2}U_0+c_0I_3,\\
	&c_0=\frac{\lambda^2}{4}-\frac{1}{2b_1b_2\lambda^2}+\frac{b_1+b_2}{b_1b_2}(a_1^2b_1+a_2^2b_2-1).
\end{align*}
Define $z=\lambda^{-2}$ and suppose $\kappa$ satisfies
\begin{align}\label{egv}
	\det[(\kappa-z)I_3-U_0]=0.
\end{align}
The eigenvector belonging to the eigenvalue $\kappa_m-z~(m=1,2,3)$ can be chosen as $\alpha_m=\left(1,\frac{a_1b_1}{\lambda(b_1+\kappa_m)},\frac{a_2b_2}{\lambda(b_2+\kappa_m)} \right)^T.$ The eigenvalue decomposition of $U_0$ shows $U_0M=ME,$ where $E=\diag(\kappa_1-z,\kappa_2-z,\kappa_3-z),~M=(\alpha_1,\alpha_2,\alpha_3).$ Then the solution of system (\ref{Laxpair2}) can be derived as
\begin{align*}
	\Phi_{0}=M\diag(\e^{\vartheta_1},\e^{\vartheta_2},\e^{\vartheta_3}),
\end{align*}
where $\vartheta_m=\ii(\kappa_m-z)\left[\tau+\frac{1}{2b_1b_2z}(\kappa_m-z+b_1+b_2)\xi\right]+c_0\xi.$
Therefore the fundamental solution of Lax pair (\ref{Laxpair1}) is given by $$\Phi=DM\diag(\e^{\vartheta_1},\e^{\vartheta_2},\e^{\vartheta_3}).$$
In addition, (\ref{egv}) gives
\begin{align}\label{6}
		\frac{\kappa}{z}-2+\frac{a_1^2b_1^2}{\kappa+b_1}+\frac{a_2^2b_2^2}{\kappa+b_2}=0.
\end{align}
From (\ref{6}) and its conjugation equation, $\kappa_m,\kappa_l$  satisfy the following equation
\begin{align}\label{7}
	\frac{\lambda^2\kappa_m-\lambda^{*2}\kappa_l^*}{\kappa_l^*-\kappa_m}+\sum_{i=1}^2\frac{a_i^2b_i^2}{(\kappa_m+b_i)(\kappa_l^*+b_i)}=0,
\end{align}
where $1\le m,l\le 3.$

Now we use DT method to find the fundamental soliton solution.
By loop group method \cite{Terng}, the general $N$-fold DT can be expressed as
$T_N=I+\sum_{i=1}^{N}\left[\frac{A_i}{\lambda-\lambda_i^*}-\frac{\sigma_3A_i\sigma_3}{\lambda+\lambda_i^*}\right]$ \cite{Ling1},
where $I$ is the identity matrix, $A_i=|x_i\rangle\langle y_i|J$, $J=\diag(1,-1,-1)$, $\langle y_i|=|y_i\rangle^\dagger$, $|y_i\rangle=DM\diag(\e^{\vartheta_1},\e^{\vartheta_2},\e^{\vartheta_3})|_{\lambda=\lambda_i}(c_1,c_2,c_3)^T\equiv(\varphi_i,\psi_i,\chi_i)^T$, 
\begin{align*}
	&[|x_{1,1}\rangle\!,\!\cdots\!,\!|x_{N,1}\rangle]\!=\![|y_{1,1}\rangle\!,\!\cdots\!,\!|y_{N,1}\rangle]B^{-1}\!,\!B \! =\!(B_{ij})_{N\!\times\! N},\\
	&[|x_{1,k}\rangle\!,\!\cdots\!,\!|x_{N,k}\rangle]\!=\![|y_{1,k}\rangle\!,\!\cdots\!,\!|y_{N,k}\rangle]H^{-1}\!,\! H \!=\!(H_{ij})_{N\!\times\! N},
\end{align*}
$k=2,3,$ and
\begin{align*}
	B_{ij}&=\frac{\langle y_i|J|y_j\rangle}{\lambda_i^*-\lambda_j}\!+\!\frac{\langle y_i|J\sigma_3|y_j\rangle}{\lambda_i^*+\lambda_j},H_{ij}=\frac{\langle y_i|J|y_j\rangle}{\lambda_i^*-\lambda_j}\!-\!\frac{\langle y_i|J\sigma_3|y_j\rangle}{\lambda_i^*+\lambda_j}.
\end{align*}
The iterative relation of the potential function satisfies
$$Q[N]=Q+\sum_{i=1}^{N}(A_i-\sigma_3A_i\sigma_3).$$
More over,
\begin{align}\label{8}
	u_s[N]= a_s\frac{~~\det H^{[s]}}{\det H}\e^{\ii\omega_s},
\end{align}
where
\begin{align*}
	H^{[s]}&=H+2u_s^{-1}Y_1^\dagger Y_{s+1},Y_1=(\varphi_1,\cdots,\varphi_N)^T,\\
	Y_2&=(\psi_1,\cdots,\psi_N)^T,Y_3=(\chi_1,\cdots,\chi_N)^T.
\end{align*}
To obtain dark/anti-dark soliton, we need to choose $c_2=c_3=0$ and $\lambda\in\ii\mathbb{R}\cup\mathbb{R}$. Denoting $\kappa_{1,i}=\kappa_1|_{\lambda=\lambda_i}$, then for $i\neq j$, combing with (\ref{7}), we obtain
\begin{subequations}\label{9}
	\begin{align}
		H_{ij}&=\frac{2\lambda_j^{-1}|c_1|^2\kappa_{1,i}^*}{\kappa_{1,j}-\kappa_{1,i}^*}\e^{\vartheta_{1,i}^*+\vartheta_{1,j}},\\
		H^{[s]}_{ij}&=\frac{2\lambda_j^{-1}|c_1|^2\kappa_{1,j}}{\kappa_{1,j}-\kappa_{1,i}^*}\frac{b_s+\kappa_{1,i}^*}{b_s+\kappa_{1,j}}\e^{\vartheta_{1,i}^*+\vartheta_{1,j}}.
	\end{align}
\end{subequations}
For $H_{ii}$ and $H_{ii}^{[s]}$, we use the aid of technique in \cite{Ling3}. When $\lambda\rightarrow\pm\lambda^*$ and $\kappa_2\rightarrow\kappa_1^*,$  we have $\vartheta_2\rightarrow-\vartheta_1^*$. Define $|g_i\rangle = |y_{i}\rangle+\gamma_i(\lambda_i^2-\lambda_i^{*2})D(\alpha_{2}\e^{\vartheta_{2}}|_{\lambda=\lambda_i})$ and  $\widehat{J}_{ij}=\frac{J(\lambda_i^*+\lambda_j)-J\sigma_3(\lambda_i^*-\lambda_j)}{\lambda_i^{*2}-\lambda_j^2}$, then for appropriate $\gamma_i$ that makes the limit exist, one obtains
\begin{align}\label{10}
	H_{ii}\equiv&\lim_{\lambda_i\rightarrow\pm\lambda_i^*,\kappa_2\rightarrow\kappa_1^*}\langle y_{i}|\widehat{J}_{ii}|g_i\rangle \nonumber\\
	=&\frac{2\lambda_i^{-1}|c_1|^2\kappa_{1,i}^*}{\kappa_{1,i}-\kappa_{1,i}^*}\e^{2\Re(\vartheta_{1,i})}+\frac{2\lambda_i^{-1}|c_1|^2|\kappa_{1,i}|}{\kappa_{1,i}-\kappa_{1,i}^*}\varepsilon_i,
\end{align}
where $\varepsilon_i\in\mathbb{R}\backslash\{0\}$ is a free parameter that regulates the spatial position of a single soliton. The coefficient $\frac{2\lambda_i^{-1}|c_1|^2|\kappa_{1,i}|}{\kappa_{1,i}-\kappa_{1,i}^*}$ is chosen to simplify the subsequent analysis. Combining with (\ref{8})-(\ref{10}), after eliminating $2^N\prod_{j=1}^{N}\lambda_j^{-1}|c_1|^2$ we obtain
\begin{align}\label{11}
	u_s[N]= a_s\frac{~~\det H^{[s]}}{\det H}\e^{\ii\omega_s},
\end{align}
where
\begin{align*}
	H^{[s]}_{ij}&=\frac{\kappa_{1,j}}{\kappa_{1,j}-\kappa_{1,i}^*}\frac{b_s+\kappa_{1,i}^*}{b_s+\kappa_{1,j}}\e^{\vartheta_{1,i}^*+\vartheta_{1,j}}+\frac{\varepsilon_i\delta_{ij}|\kappa_{1,i}|}{\kappa_{1,i}-\kappa_{1,i}^*},\\
	H_{ij}&=\frac{\kappa_{1,i}^*}{\kappa_{1,j}-\kappa_{1,i}^*}\e^{\vartheta_{1,i}^*+\vartheta_{1,j}}+\frac{\varepsilon_i\delta_{ij}|\kappa_{1,i}|}{\kappa_{1,i}-\kappa_{1,i}^*}.
\end{align*}
Here $\delta_{ij}$ is the Kronecker delta and $\varepsilon_i$ is a real parameter controlling the soliton's spatial position.

\section{The condition of MHMVS}\label{App_k}
This section gives the detail to obtain Eq. \eqref{16} and Eq. \eqref{18}. Combining Eq. (\ref{spectrumcondition}) and its conjugation, Cramer's rule shows
\begin{subequations}\label{a1a2eq}
	\begin{align}
		a_1^2&=\frac{[\Re(\kappa)+b_1]^2+[\Im(\kappa)]^2}{zb_1^2(b_1-b_2)}\left(2z-2{\Re(\kappa)}-b_2\right),\\
		a_2^2&=\frac{[\Re(\kappa)+b_2]^2+[\Im(\kappa)]^2}{zb_2^2(b_2-b_1)}\left(2z-2{\Re(\kappa)}-b_1\right).
	\end{align}
\end{subequations}
Note that in the DT, the background amplitudes are consistent. Thus for arbitrary two different spectral parameters $z_p,z_q$ and the associated $\kappa_p,\kappa_q$, $a_s^2(z_p,\kappa_p)=a_s^2(z_q,\kappa_q)$ holds, which reads
\begin{subequations}\label{15}
	\begin{align}
		[\Im(\kappa_p)]^2\mu_{p,2}-[\Im(\kappa_q)]^2\mu_{q,2}&=f_1,\\
		[\Im(\kappa_p)]^2\mu_{p,1}-[\Im(\kappa_q)]^2\mu_{q,1}&=f_2
	\end{align}
\end{subequations}
after inserting Eq. (\ref{14}) into Eqs. (\ref{a1a2eq}), where $\mu_{p,k}=z_p^{-1}(2z_p-2\Re(\kappa_p)-b_k),$ and $f_1=[\Re(\kappa_q)+b_1]^2\mu_{q,2}-[\Re(\kappa_p)+b_1]^2\mu_{p,2},f_2=[\Re(\kappa_q)+b_2]^2\mu_{q,1}-[\Re(\kappa_p)+b_2]^2\mu_{p,1}.$  When $v_p=v_q,$ calculation gives
\begin{align*}
	\det\begin{pmatrix}
		\mu_{p,2} & -\mu_{q,2}\\
		\mu_{p,1} & -\mu_{q,1}
	\end{pmatrix}=2v_pb_1b_2\Delta b(z_q^{-1}-z_p^{-1}),
\end{align*}
where $\Delta b=\frac{b_1-b_2}{2}.$ Hence if $v_p\neq 0$, Eq. (\ref{15}) give the unique solution \eqref{16}.
Now, by substituting the parameters determined by Eqs. (\ref{a1a2eq}), Eq. (\ref{14}) and Eq. (\ref{16}) into Eq. (\ref{solution}), one obtains a MHMVS corresponding to the spectral parameters $z_p$ and $z_q$ with velocity $v_p=v_q\neq 0$. On the other hand, for the case $v_p=v_q=0$, the augmented matrix of Eqs. (\ref{15}) can
be rewritten as
\begin{align}\label{17}
	\begin{pmatrix}
		\mu_{p,2} & -\mu_{q,2} & f_1\\
		\mu_{p,1} & -\mu_{q,1} & f_2
	\end{pmatrix}=\begin{pmatrix}
		b_1 \\
		b_2
	\end{pmatrix}\begin{pmatrix}
		z_p^{-1} & -z_q^{-1} & f
	\end{pmatrix},
\end{align}
where $f=z_q-z_p+(\Delta b)^2(z_q^{-1}-z_p^{-1}).$
A direct calculation yields \eqref{18}.

\section{The exact expressions} \label{App2}
The explicit expressions of $g_s,f_s$ in Eq. \eqref{dp} for the parameter choice in Fig. \ref{fig2} are given by
\begin{align*}		
	&f_1=\\
	&-125x^6 \left(10 \sqrt{2} \varpi ^4+3 \sqrt{5} \varpi ^3\right)\\
	&-250 x^5\varpi^2 \left(15 \varpi ^2+8 \sqrt{10} \varpi +7 \right)\\
	&-25 x^4\varpi \left(50 \sqrt{2}
	\varpi ^3+128 \sqrt{5} \varpi ^2+168 \sqrt{2} \varpi +15 \sqrt{5}  \right)\\
	&-20x^3\varpi \left(30 \sqrt{10} \varpi ^2+211 \varpi +30 \sqrt{10}
	\right)\\
	&-5x^2 \left(15 \sqrt{5} \varpi ^3+168 \sqrt{2} \varpi ^2+128 \sqrt{5} \varpi +50 \sqrt{2}\right)\\
	&-10x \left(7 \varpi ^2+8 \sqrt{10} \varpi +15\right)-(3
	\sqrt{5} \varpi +10 \sqrt{2}),\\
	&f_2=\\
	&+125 \sqrt{5} x^6 \varpi ^3+50x^5\varpi^2 \left(25 \varpi ^2+8 \sqrt{10} \varpi +15 \right)\\
	&+25x^4\varpi \left(24 \sqrt{5} \varpi ^2+20 \sqrt{2} \varpi +5
	\sqrt{5}  \right)+580 x^3 \varpi ^2\\
	&+5x^2\varpi \left(5 \sqrt{5} \varpi ^2+20 \sqrt{2} \varpi +24 \sqrt{5}  \right)\\
	&+2x \left(15\varpi ^2+8 \sqrt{10}
	\varpi +25\right)+\sqrt{5} \varpi,\\
	&g_1=\\
	&+3125 x^8 \varpi ^4+750 x^7 \varpi ^3 \left(10 \sqrt{2} \varpi +7 \sqrt{5}\right)\\
	&+50 x^6 \varpi ^2 \left(300 \varpi ^2+212 \sqrt{10} \varpi +345\right)\\
	&+50 x^5
	\varpi  \left(150 \sqrt{2} \varpi ^3+332 \sqrt{5} \varpi ^2+544 \sqrt{2} \varpi +105 \sqrt{5}\right)\\
	&+5 x^4 \left(625 \varpi ^4+1120 \sqrt{10} \varpi ^3+5888
	\varpi ^2+1120 \sqrt{10} \varpi +625\right)\\
	&+10 x^3 \left(105 \sqrt{5} \varpi ^3+544 \sqrt{2} \varpi ^2+332 \sqrt{5} \varpi +150 \sqrt{2}\right)\\
	&+x^2 \left(690
	\varpi ^2+424 \sqrt{10} \varpi +600\right)\\
	&+6 x \left(7 \sqrt{5} \varpi +10 \sqrt{2}\right)+5,
\end{align*}
\newpage
\begin{align*}	
	&g_2=\\
	&+625 x^8 \varpi ^4+750 \sqrt{5} x^7 \varpi ^3+50x^6 \left(4 \sqrt{10} \varpi ^3+45 \varpi ^2\right)\\
	&+50x^5\varpi \left(4 \sqrt{5} \varpi^2 +20 \sqrt{2} \varpi +15
	\sqrt{5} \right)
\end{align*}
\newpage
\begin{align*}		
	&+5x^4 \left(125 \varpi ^4+80 \sqrt{10} \varpi ^3+304 \varpi ^2+80 \sqrt{10} \varpi +125\right)\\
	&+10x^3\varpi \left(15 \sqrt{5} \varpi ^2+20
	\sqrt{2} \varpi +4 \sqrt{5}  \right)\\
	&+2x^2 \left(45 \varpi ^2+4 \sqrt{10} \varpi \right)+6 \sqrt{5} x \varpi +1.
\end{align*}

\end{document}